\documentclass{fairmeta}

\usepackage[table,xcdraw]{xcolor}
\usepackage{graphics}
\usepackage{graphicx} 
\usepackage{subcaption} 
\usepackage{multirow}
\usepackage{amssymb}  
\usepackage{booktabs}
\usepackage{pgfplots}
\usepackage{comment}
\pgfplotsset{compat=1.17}
\usetikzlibrary{patterns}

\definecolor{tabblue}{HTML}{1F77B4}
\definecolor{tabbluelight}{HTML}{AEC7E8}
\definecolor{taborange}{HTML}{FF7F0E}
\definecolor{taborangelight}{HTML}{FFBB78}
\definecolor{tabgreen}{HTML}{2CA02C}
\definecolor{tabgreenlight}{HTML}{98DF8A}
\definecolor{tabred}{HTML}{D62728}
\definecolor{tabredlight}{HTML}{FF9896}
\colorlet{fsspecmean}{tabblue}
\colorlet{fsspecp99}{tabbluelight}
\colorlet{nativemean}{taborange}
\colorlet{nativep99}{taborangelight}

\usepackage{listings}
\newcommand{\sysname}{PRISM}

\definecolor{redColor}{HTML}{1F77B4} 
\definecolor{blueColor}{HTML}{D62728}   
\definecolor{greenColor}{HTML}{2CA02C} 

\title{Deployed-System Paper: \sysname{}: Evaluating POSIX Storage Systems for AI Research Workflows}

\author[1,*]{Adithya Kumar}
\author[2]{Abhinandan Prativadi}
\author[2]{Aditya Basu}
\author[2]{Jacob Kahn}
\author[2]{Parth Malani}
\author[2]{Leo Huang}
\author[2]{Kalyan Saladi}

\affiliation[1]{Indian Institute of Technology, Gandhinagar}
\affiliation[2]{FAIR at Meta}
\affiliation[*]{\em This work was done while the author was at FAIR.}

\abstract{Large GPU clusters for AI research rely on POSIX-based storage spanning code authoring, data preparation, and model checkpointing. The metric that matters is researcher iteration speed, not just peak throughput -- but current benchmarks target the latter: synthetic tools stress peak bandwidth and IOPS, emulators replay captured training I/O, and neither execute real framework operations or captures the bursty, heterogeneous, metadata-heavy patterns of research.

We present \sysname{}, an extensible benchmark suite that executes real workloads across key stages of AI research including version control, environment setup, data preparation, data loading, model checkpointing, and synthetic data generation.
\sysname{} exercises what researchers actually run and thereby serves not only as a benchmark but also as a qualification gate for new storage systems before they are consumed by researchers.

We share our experience using \sysname{} for over 18 months in AI research clusters with thousands of Nvidia hopper class GPUs consuming petabyte scale Lustre and NFS based storage systems in characterizing research workloads, identifying the appropriate system for different cluster environments and detecting performance regressions.
As an example, \sysname{} surfaced an \texttt{openat} lock contention pathology that caused an 8$\times$ slowdown under Fully Sharded Data Parallel (FSDP) checkpointing which went uncaught in emulated benchmarks.
We plan to open source \sysname{} as a practical tool for selecting and validating storage for AI research infrastructure.
}

\date{\today}
\correspondence{Kalyan Saladi at \email{skalyan@meta.com}}

\begin{document}

\maketitle

\section{Introduction}
\textbf{AI Research Clusters \& Storage:}
The race towards superintelligence is driving multi-billion dollar investments in AI compute clusters centered on acquiring and efficiently operating the latest GPUs. Leading AI labs compete for this compute, whether by building in-house infrastructure or using GPU neo-clouds.

While GPUs dominate attention, keeping them utilized depends critically on storage. Neo-clouds that race to offer the latest GPUs are rarely held to the same standard for storage, shifting the burden of achieving high GPU utilization to researchers and infrastructure teams. AI research productivity therefore requires infrastructure that maximizes flexibility and minimizes the impediments to researchers.

This demands seamless integration of compute and data across a rapidly evolving set of tools and frameworks for prototyping, training, and refining models. Storage is central to this stack: it directly impacts data processing speed, training time, and end to end cluster efficiency while keeping researcher velocity in mind.

Suboptimal storage can have an enormous effect. As we show later, suboptimal usage of a storage system on a major cloud vendor causes a 3$\times$ slowdown for a common data loading job (\S~\ref{sec:lustre_vs_nas}), and a critical bug in a major storage vendor's system caused an 8$\times$ increase in checkpoint latency and a dramatic drop in cluster efficiency (\S~\ref{sec:regression}).

\textbf{Role of storage in AI research workflows:}
AI research workflows differ fundamentally from stable production training. They are dynamic, fragmented, interactive, multimodal, and experimental: rapidly reshaped, repetitive in the short term, and often discarded in the long term. This requires high adaptability from the underlying infrastructure.

A typical lifecycle proceeds from hypothesis prototyping to dataset pre-processing, small-scale pre-training, large scale training, post-training and fine-tuning, and model publication.

In this fluid environment, POSIX based systems such as NFS and Lustre are critical for productivity, especially during early prototyping. While large scale pre-training of frontier models may require bespoke storage for peak performance, battle tested POSIX systems provide a compatibility layer that ensures interoperability across a vast ecosystem of off the shelf and custom tools.

This role is evident across four stages: (i) environment management (e.g., cloning repositories, building packages and conda environments, handling configuration archives), (ii) data generation (e.g., downloading datasets, synthetic data generation), (iii) data curation and consumption (e.g., data loading for pre-training, dataset validation, pre-processing), and (iv) checkpoint management.

This dependency means that the storage system must not become a bottleneck. The scrappy, unpredictable nature of research produces access patterns that are often unforeseen and, in our experience, can lead to significant time lost debugging what turn out to be dead end paths.

\textbf{Evaluating POSIX storage with \sysname{}:}
Storage system bottlenecks directly reduce GPU utilization, making their evaluation critical given the cost of modern GPU clusters. 
Yet existing widely known benchmark suites fall short: general purpose benchmarks like fio~\cite{fio} and elbencho~\cite{elbencho} measure bandwidth and IOPS, while application specific suites like MLPerf Storage~\cite{mlperfstorage} emulate rather than execute training I/O, neither captures the scope and dynamicity of real research, where heavy pre-training (e.g., data preprocessing, synthetic data generation) competes unpredictably with interactive research workflows (e.g., version control).

To close this gap, we developed \sysname{}\footnote{\sysname{}: \underline{P}OSIX \underline{R}esearch \underline{I}nfrastructure \underline{S}torage \underline{M}easurement System}, a canonical set of realistic applications modeling workflows observed on our clusters over 18 months. \sysname{} is designed to (i) measure application level performance rather than low level primitives, (ii) extend to diverse Python based applications replicating complex access patterns, and (iii) scale from single node validation to medium scale GPU clusters (up to 1K GPUs).

We continuously use it to (i) qualify new storage vendors with large GPU deployments, (ii) compare competing storage offerings on public clouds, and (iii) evaluate vendor and kernel upgrades across our fleet.

\textbf{Key Contributions:}
\begin{enumerate}
\itemsep0em

\item \textbf{Characterize use of POSIX storage in AI research clusters:} We analyze 90 days of telemetry from three clusters, each with more than 8K Nvidia Hopper class GPUs, spanning petabytes of managed Lustre, vendor NAS appliances, and software-defined NAS. We show that research storage is metadata-dominated with volatile capacity usage. (Section\S~\ref{sec:posix}).

\item \textbf{\sysname{}, a benchmark suite that executes rather than
emulates:} \sysname{} runs the real research workflow operations, e.g., PyTorch
Distributed Data Parallel (DDP) and Fully Sharded Data Parallel (FSDP) checkpointing, distributed data loading, and version control, directly on the cluster with support for seamlessly adding other Python workloads as plugins.
(Section\S~\ref{sec:fsbench}).

\item \textbf{Share deployment experience and insights:}
We report 18 months of operating \sysname{} on production AI research
clusters across five use cases (see \S~\ref{sec:using}).
\sysname{} caught a vendor NFS regression in
which \texttt{openat} stalled for $\approx$5\,s under concurrent access,
inflating FSDP checkpoint-restore latency by up to 8$\times$, which the
vendor confirmed and patched. Such bugs can be trivially surfaced by \sysname{}
that runs actual workloads instead of replaying I/O traces.
We further show that no single POSIX
architecture dominates, Lustre wins on large-file checkpoint writes while
the NAS appliance is up to 80$\times$ faster on metadata at 256 clients, and
that the \texttt{fsspec} abstraction layer imposes non-trivial overhead on
metadata heavy paths.

\item \textbf{Leverage \sysname{} for vendor evaluations:}
We were able to leverage \sysname{} to evaluate NAS offerings from several storage vendors and draw out the actual performance characteristics that meaningfully impacts the researcher experience, e.g., per-operation metadata latency under concurrent clients, checkpoint-restore behavior at the straggler rank, and the tail latency of interactive operations such as environment activation.

\end{enumerate}

\section{Storage Systems for AI Research Workflows}
\label{sec:posix}

\begin{table}[]
\resizebox{\linewidth}{!}{%
\begin{tabular}{r|cccc}
\hline
\textbf{Storage offering} &
  \multicolumn{1}{l}{\textbf{Compatibility}} &
  \multicolumn{1}{l}{\textbf{Deployability}} &
  \multicolumn{1}{l}{\textbf{Configurability}} &
  \multicolumn{1}{l}{\textbf{Scalability}} \\ \hline
\begin{tabular}[c]{@{}r@{}}POSIX filesystems\\ (NFS, Lustre)\end{tabular} & High   & Easy & Moderate & Medium \\ \hline
\begin{tabular}[c]{@{}r@{}}Object store\\ (S3, GCS)\end{tabular}          & Medium & Easy & Low      & High   \\ \hline
\begin{tabular}[c]{@{}r@{}}Distributed systems\\ (HDFS, Ceph)\end{tabular} & Low    & Hard & High     & High   \\ \hline
\end{tabular}
}

\caption{Tradeoffs between different cloud storage offerings for AI research workflows}
\label{tab:cloud-storage-tradeoffs}
\end{table}
Table~\ref{tab:cloud-storage-tradeoffs} compares cloud storage offerings along four dimensions: compatibility, deployability, configurability, and scalability.

Cloud-native object stores (e.g., Amazon S3, Google GCS) offer trivial deployment and near-unbounded scalability through simple HTTP APIs, making them the de facto choice for dataset and artifact archival. They are, however, ill-suited to interactive use: access latency is high, consistency guarantees are weak, and their non-POSIX APIs compel either application rewrites or intermediary shims (e.g., s3fs, specialized libraries) for general-purpose consumption.

Distributed systems such as HDFS and Ceph provide scalable, fault-tolerant storage with deep integration into big-data frameworks, but trade general file semantics for performance and remain operationally complex to deploy and configure on the public cloud.

In contrast, POSIX-compliant systems---NFS (e.g., AWS EFS, FSx for ONTAP, FSx for OpenZFS) and Lustre (e.g., FSx for Lustre, Azure Managed Lustre)---expose the conventional file-and-directory abstraction, rendering remote storage indistinguishable from local directories with well-defined consistency semantics. This ubiquity maximizes compatibility with existing tools, users, and, increasingly, agents, and drastically simplifies integration with established software, scripts, and workflows.

We observe that POSIX storage yields outsized gains in research productivity: the filesystem abstraction assumed by the open-source ecosystem (e.g., PyTorch, NumPy, conda) lowers the barrier to experimentation and streamlines operational tasks such as scripting and debugging, establishing POSIX as the preferred default for interactive work and training runs.

Nonetheless, while managed POSIX offerings (e.g., FSx) simplify deployment, selecting and tuning the appropriate variant---e.g., high-throughput Lustre versus simpler NFS---to meet the performance demands of AI research workloads without undue operational complexity remains an open challenge.

\subsection{fsspec}
\label{sec:fsspec_background}
\texttt{fsspec} is a Python library that provides a unified interface across filesystem backends, and has gained traction in AI research workflows through its adoption by data-loading libraries. 
When training deep learning models, researchers access data spread across local disks, cloud object stores (e.g., S3, GCS), HDFS clusters, and HTTP endpoints. Rather than writing backend-specific I/O code, libraries built on \texttt{fsspec} expose these locations through a consistent \texttt{protocol://path} URL syntax, so the same loader code can target local or remote data by changing only the URL scheme. 
PyTorch-adjacent libraries such as \texttt{torchdata}, Hugging Face's \texttt{datasets}, and PyTorch Lightning build on \texttt{fsspec} to read training data and to write checkpoints directly to cloud storage without intermediate local copies (e.g., via \texttt{fsspec}-backed file-like objects passed to \texttt{torch.save}).

However, \texttt{fsspec} addresses only programmatic data access; the operations surrounding the training loop still assume POSIX semantics. Package managers such as \texttt{conda} and \texttt{uv} rely on hard links, file locking, and permissions when installing environments and populating package caches. Researchers orchestrate experiments with shell scripts that chain POSIX-only utilities (\texttt{ls}, \texttt{find}, \texttt{grep}, \texttt{rsync}, \texttt{awk}, \texttt{sed}, \texttt{jq}), and debug by tailing logs (\texttt{tail -f}), paging through outputs (\texttt{less}), and diffing configuration files, often through editors and notebook interfaces that expect local paths. Monitoring and profiling tools similarly assume filesystem primitives such as atomic rename for checkpoint writes and file-watching (\texttt{inotify}) for live log updates, neither of which object stores provide natively. Consequently, even when training data resides in cloud storage and is accessed via \texttt{fsspec}, researchers typically mount shared NFS volumes for everything else: code repositories, virtual environments, experiment tracking, intermediate outputs, and the collaborative workspace where team members browse each other's results.

\subsection{Storage Infrastructure for Production vs.\ Research AI Workflows}

\begin{table}[t]
\resizebox{\linewidth}{!}{%
\begin{tabular}{|l|c|c|}
\hline
\textbf{Dimension} &
  \textbf{Production AI training} &
  \textbf{Research AI training} \\ \hline
\textbf{High-level goals} &
  \begin{tabular}[c]{@{}c@{}}Repeatability, \\ Efficiency, \\ Strict SLOs\end{tabular} &
  \begin{tabular}[c]{@{}c@{}}Flexibility, \\ Interactivity, \\ Low friction\end{tabular} \\ \hline
\textbf{Workflow} &
  Well-defined, repeatable &
  Rapid, ad hoc experiments \\ \hline
\textbf{Performance target} &
  \begin{tabular}[c]{@{}c@{}}Predictable access patterns, \\ high throughput\end{tabular} &
  \begin{tabular}[c]{@{}c@{}}Mixed access patterns (sequential/random), \\ low latency\end{tabular} \\ \hline
\textbf{Resource demand} &
  Stable, predictable volume &
  Highly bursty, irregular \\ \hline
\textbf{Scaling} &
  \begin{tabular}[c]{@{}c@{}}Right-sized, \\ high utilization\end{tabular} &
  \begin{tabular}[c]{@{}c@{}}Easy to grow and reconfigure, \\ rebalance with minimal coordination\end{tabular} \\ \hline
\textbf{Data lifecycle} &
  \begin{tabular}[c]{@{}c@{}}Governed, strict schemas, \\ immutable datasets, lineage tracking\end{tabular} &
  \begin{tabular}[c]{@{}c@{}}Intermediate artifacts in scratch space, \\ frequent updates\end{tabular} \\ \hline
\textbf{Data security standards} &
  \begin{tabular}[c]{@{}c@{}}Sensitive first-party data, \\ strict access control\end{tabular} &
  \begin{tabular}[c]{@{}c@{}}Open third-party data, \\ low barrier to access\end{tabular} \\ \hline
\end{tabular}
}
\caption{High-level differences in how production and research AI training use storage infrastructure.}
\label{tab:prod-res-ai-differences}
\end{table}

The storage requirements of research AI training diverge from those of production AI training along several dimensions. Production training is engineered for \textit{repeatability, efficiency, and strict service-level objectives}; research training prioritizes \textit{flexibility, interactivity, and low friction} for exploratory work. Table~\ref{tab:prod-res-ai-differences} summarizes the distinction; we discuss the salient points below.

Production training runs in well-defined, automated pipelines on predictable schedules, ingesting stable data volumes under governed lifecycles: strict schemas, immutable datasets, and strong lineage tracking, with sensitive first-party data under tight access control. This regularity favors specialized high-performance object stores with strong durability and availability guarantees, and permits ``right-sizing'' of resources with bespoke, performance-tuned APIs that trade POSIX compliance for throughput.

Research training, in contrast, supports ad hoc experimentation over heterogeneous workloads: bursty I/O demand, mixed sequential and random access, and frequent updates to intermediate artifacts and scratch space, yielding many small, short-lived files. An ever-evolving toolchain and frictionless collaboration favor generic POSIX-based storage (e.g., NFS, Lustre) as the common denominator for open-source tools. Such storage must be easy to grow, reconfigure, and rebalance with minimal coordination, and researchers typically work with open third-party data under minimal access barriers.

This divergence calls for a rethinking of storage performance analysis. Traditional benchmarks~\cite{fio,elbencho,filebench} are conventionally configured around steady-state, sequential, large-file access patterns, which do not capture the bursty demand, mixed random/sequential I/O, and small-file dominance of research environments. Characterizing research storage therefore requires benchmarks tuned to these highly variable access patterns. This paper focuses on storage for research AI training, where POSIX-based systems play a central role: we characterize how research workloads exercise the underlying storage and how benchmarking should be performed for such environments.

\subsection{Characterizing POSIX Storage in Our Research Environments}
\begin{figure*}[t]
    \centering
    \begin{subfigure}{0.24\textwidth}
        \includegraphics[width=\textwidth]{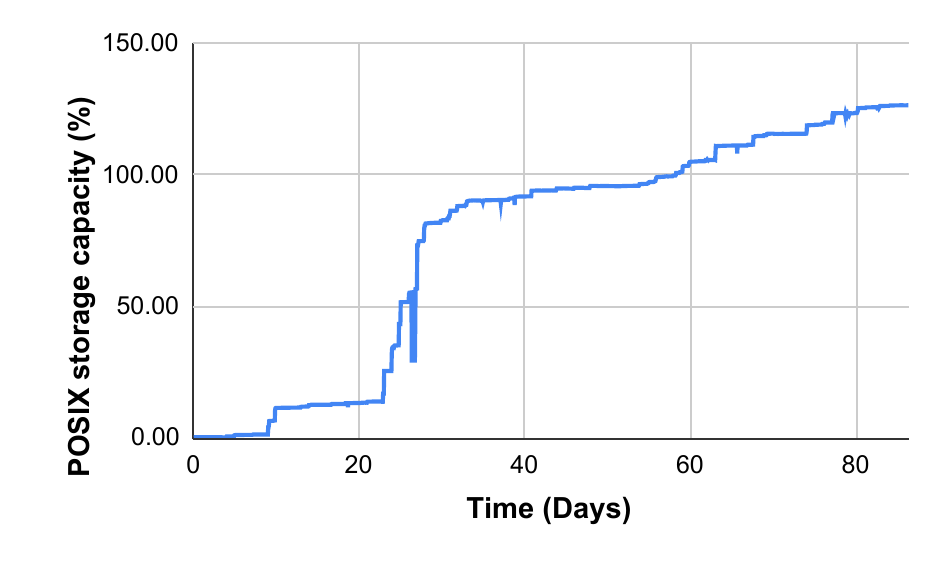}
        \caption{Capacity growth over 90 days (\texttt{Lustre})}
        \label{fig:cap_lustre}
    \end{subfigure}
    \begin{subfigure}{0.24\textwidth}
        \includegraphics[width=\textwidth]{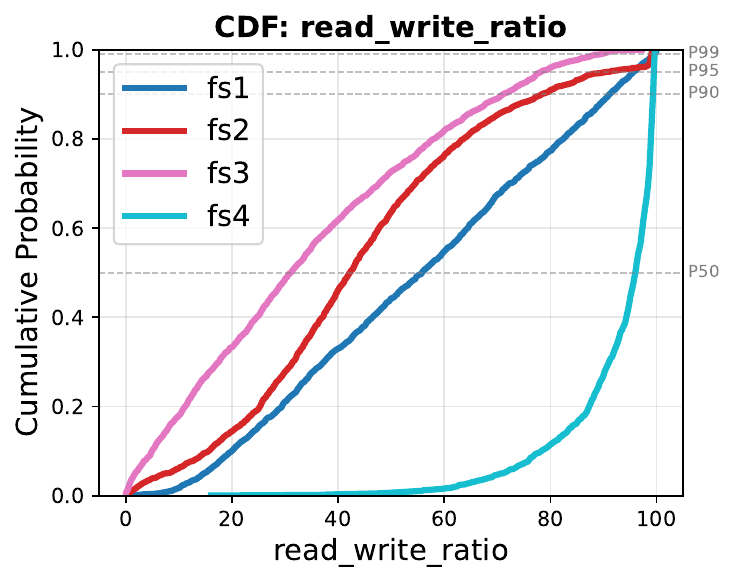}
        \caption{Read/write throughput (\texttt{NAS})}
        \label{fig:throughput_nas}
    \end{subfigure}
    \begin{subfigure}{0.24\textwidth}
        \includegraphics[width=\textwidth]{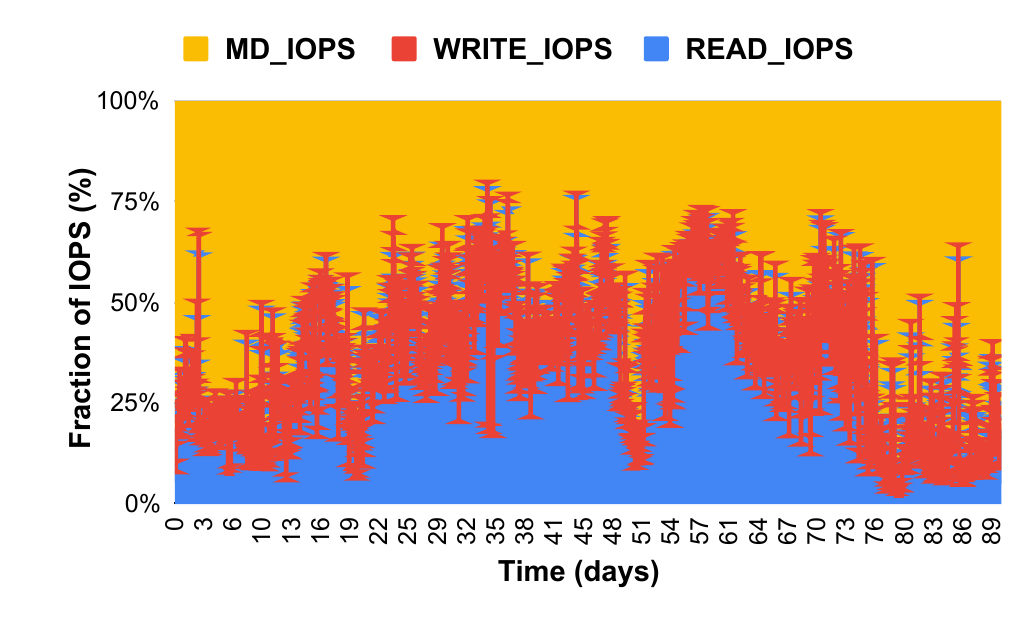}
        \caption{Read, write, and metadata IOPS (\texttt{NAS})}
        \label{fig:md_nas}
    \end{subfigure}
    \begin{subfigure}{0.24\textwidth}
        \includegraphics[width=\textwidth]{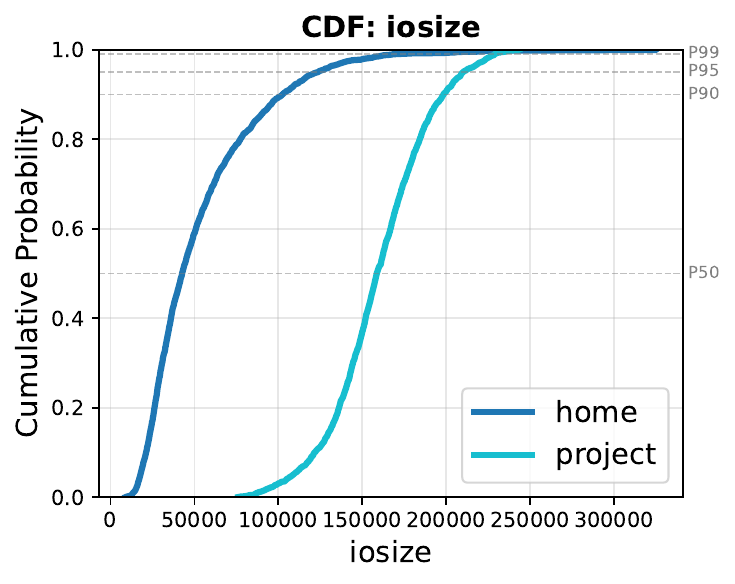}
        \caption{Average I/O size, \texttt{home} vs.\ \texttt{project} (\texttt{NAS})}
        \label{fig:blk_size_nas}
    \end{subfigure}
    \caption{Characterizing POSIX storage in research environments.}
    \label{fig:posix-characterization}
\end{figure*}

To understand the role of POSIX-based storage in research clusters, we characterize the capacity footprint and usage patterns of the primary working storage used for day-to-day experimentation, code development, and training. Each of the three clusters we study has at least 8K NVIDIA Hopper-class GPUs, and researchers rely on POSIX-based storage for home directories (code, conda environments) and shared project directories (configurations, checkpoints, datasets, logs).

We consider three types of POSIX-based storage across these clusters: (i) \textbf{Lustre (\texttt{L})} on cloud, a fully managed Lustre offering owned and operated by the cloud vendor (e.g., AWS FSx for Lustre, Azure Managed Lustre); (ii) \textbf{NAS (\texttt{NAS})} on cloud, an off-the-shelf network-attached storage appliance purchased from a storage vendor and deployed on cloud infrastructure, with hardware and software designed and managed by the vendor (e.g., NetApp, VAST, Pure Storage); and (iii) \textbf{hybrid NAS (\texttt{NAS-SW})} on cloud, a software-defined aggregated NAS solution (e.g., Hammerspace, Weka).

While the Lustre offering is fully POSIX-compliant, the NAS offerings expose only NFS (v3/v4) client APIs with limited POSIX compliance. Unlike traditional HPC settings that depend on Lustre's full POSIX semantics, we find NFS semantics sufficient for our AI research use cases in practice.

Below we present the salient characteristics of these systems and the implications that shaped our storage qualification benchmark.

\subsubsection{Capacity footprint and usage}
Figure~\ref{fig:cap_lustre} shows the capacity of the \texttt{Lustre} filesystem on one cluster over 90 days in Q4 2025. Capacity grew by over 125\% in that window, with the largest increase concentrated in a two-week span.

\textbf{Implication:} POSIX capacity is business-critical yet grows unpredictably, so qualification must capture multi-tenant performance under rapid expansion.

\subsubsection{Read/write throughput patterns}
Figure~\ref{fig:throughput_nas} shows 90-day read/write bandwidth on one \texttt{NAS} server with four filesystems and approximately 150\,GB/s of peak bandwidth. Reads dominate as expected, but there are intervals in which read and write bandwidth are nearly equal, consistent with bulk dataset ingestion occurring alongside checkpoint writes.

\textbf{Implication:} Although reads dominate on average, shared research filesystems see transient ingestion-driven write bursts that contend with read traffic. Benchmarks must reproduce this contention rather than assuming a steady read-heavy mix.

\subsubsection{Metadata IOPS}
Figure~\ref{fig:md_nas} breaks IOPS on the same server into read, write, and metadata operations. Metadata operations accounted for the majority of total IOPS, a far larger share than in typical production or HPC profiles.

\textbf{Implication:} Metadata performance must be explicitly measured and validated rather than treated as a second-order effect.

\subsubsection{I/O request sizes}
We provision two filesystem classes: \texttt{home} filesystems for code and conda environments, and \texttt{project} filesystems for checkpoints, configurations, datasets, and logs. Figure~\ref{fig:blk_size_nas} compares the average I/O size on one filesystem of each class: \texttt{home} serves much smaller I/Os driven by code and library loads, while \texttt{project} serves large I/Os.

\textbf{Implication:} The two classes have sharply different access characteristics, so qualification requires distinct small-I/O and large-I/O benchmark scenarios.

Given the centrality of POSIX storage to AI research workflows, our NFS- and Lustre-based systems must remain stable, reliable, and performant across a diverse set of vendor and cloud offerings. Our goal is a compact set of benchmark workloads that faithfully captures the characteristics above and qualifies these systems against realistic AI workload behavior before they reach users.

\section{\sysname{}: A Benchmark Framework for AI Research Clusters}
\label{sec:fsbench}

\begin{figure*}[ht]
    \centering
    \begin{subfigure}[][][t]{0.45\textwidth}
        \centering
        \includegraphics[width=\textwidth]{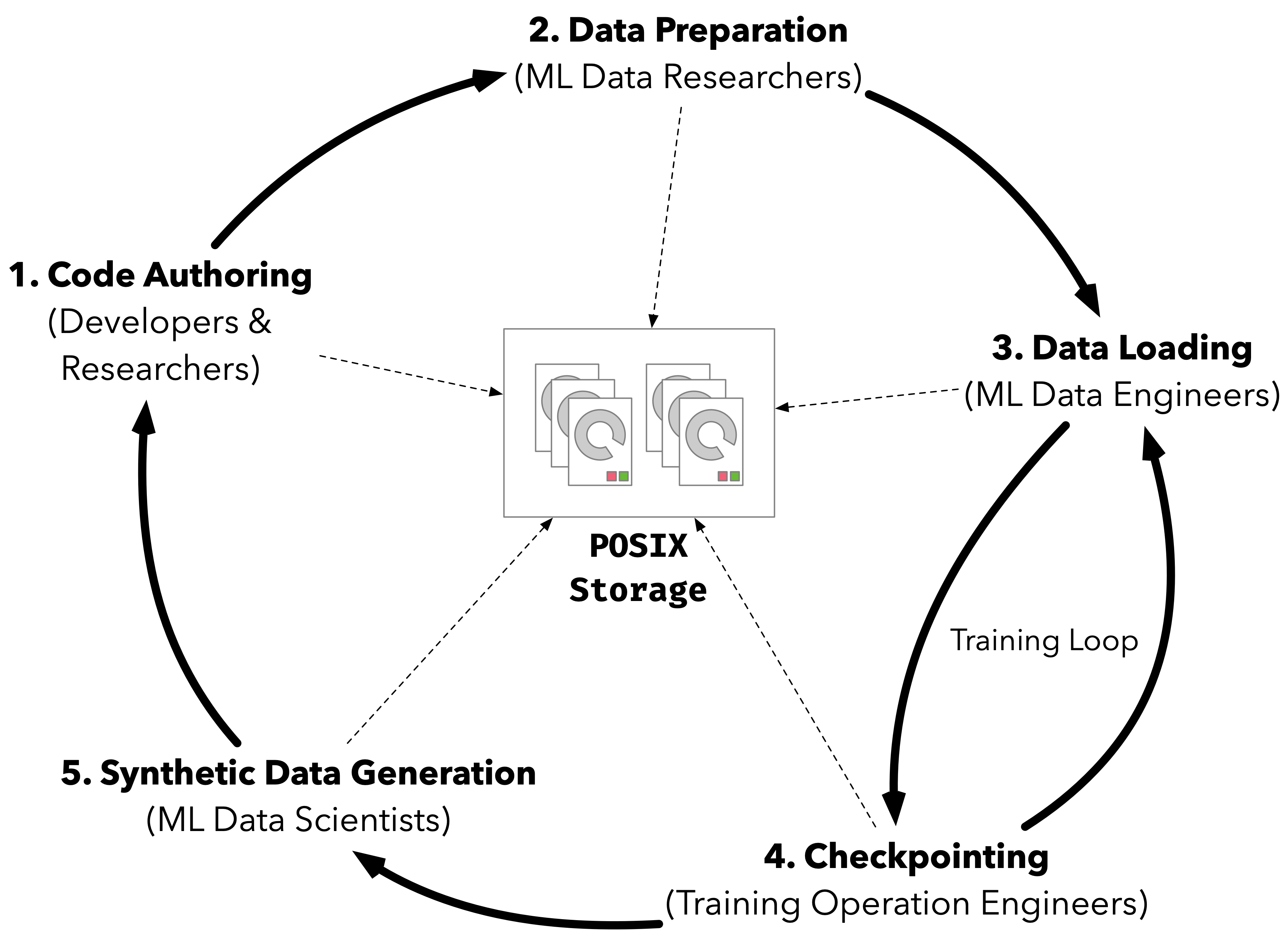}
    \end{subfigure}
    \hfill
    \begin{subfigure}{0.52\textwidth}
        \centering
        \vspace{2em}
        \resizebox{\textwidth}{!}{%
        \small
        \begin{tabular}{@{}lll@{}}
            \toprule
            \textbf{Stage} & \textbf{Activity} & \textbf{POSIX Feature} \\
            \midrule
            1. Code Authoring & \begin{tabular}[t]{@{}l@{}}Scripts, Configs, Python\\Notebooks, Conda Envs.\end{tabular} & Small file access \\[0.5em]
            2. Data Preparation & \begin{tabular}[t]{@{}l@{}}Original/Transformed\\datasets\end{tabular} & \begin{tabular}[t]{@{}l@{}}Direct I/O, Arbitrary Layout,\\Read-write support\end{tabular} \\[0.5em]
            3. Data Loading & Processed Datasets & \begin{tabular}[t]{@{}l@{}}Iterative and Random\\access\end{tabular} \\[0.5em]
            4. Checkpointing & \begin{tabular}[t]{@{}l@{}}Model weights,\\Checkpoints, Logs\end{tabular} & Fault tolerance, Streaming \\[0.5em]
            5. Synthetic Data Gen. & Generated datasets & Easy discovery by teams \\
            \bottomrule
        \end{tabular}%
        }
    \end{subfigure}
    \caption{POSIX Storage plays a central role for AI Research Workflows. The workflow diagram illustrates how various stages depend on a shared storage system, and the table summarizes the activity and POSIX features exercised at each stage.}
    \label{fig:lifecycle}
\end{figure*}

\sysname{} is a benchmark and validation framework for shared and distributed POSIX-based filesystems in AI research infrastructure.
We designed \sysname{} around two primary goals: (i) rapid validation of storage deployments when bringing up new AI clusters, and (ii) systematic comparison of performance characteristics across cloud providers and storage backends.
Unlike production clusters with well-understood workloads, research settings exhibit highly dynamic and interactive access patterns, requiring benchmarks that characterize both typical behavior and system limits.

\subsection{Benchmark Taxonomy}

We organize the \sysname{} benchmarks into five categories that follow the researcher lifecycle in Figure~\ref{fig:lifecycle}, from environment setup through data preparation and training. The set mixes application-driven benchmarks that reproduce realistic end-to-end operations (e.g., \texttt{git\_clone}, \texttt{ddp\_save}) with synthetic microbenchmarks that isolate individual operations for controlled parameter sweeps (e.g., \texttt{create\_files}, \texttt{md5\_check}). Together they span the two regimes identified in our characterization: metadata-heavy small-I/O activity and bulk large-I/O transfers. Table~\ref{tab:fsbench-benchmarks} summarizes the suite.

\subsubsection{Research Development Environments}

Reproducible research depends on consistent development environments and version-controlled codebases. The \texttt{git\_clone} benchmark measures repository cloning (directory traversal, small-file creation, permission handling), and \texttt{git\_status} measures working-tree status checks (stat-intensive directory scans and index comparisons). The \texttt{conda\_create} benchmark measures environment creation from a declarative specification, covering package extraction, bulk small-file creation, hard-link deduplication across environments, and lock-protected metadata updates; \texttt{conda\_activate} measures activation latency, dominated by script sourcing and environment-directory scans. These workloads sit in the small-I/O regime of our characterization and are latency-sensitive, stalling the researcher directly. The conda benchmarks additionally exercise POSIX-specific semantics, hard links and file locking, that distinguish storage backends for interactive use.

\subsubsection{Data Preparation and Management}
The file-operation benchmarks (\texttt{create\_files}, \texttt{list\_files}, \texttt{move\_files}, \texttt{delete\_files}) isolate the metadata-intensive operations of data processing pipelines, where researchers reorganize, filter, and transform datasets. The \texttt{folder\_bench} benchmark creates directory hierarchies of configurable depth and breadth, stressing metadata performance on nested structures typical of research data repositories. Together these target the metadata IOPS dominance observed in Figure~\ref{fig:md_nas}, while allowing controlled sweeps over operation mix and namespace shape.

\subsubsection{Data Loading}
Training data loading differs from traditional sequential file access: reads are non-sequential due to stochastic sampling yet deterministic given a seed, the same data is re-read each epoch in a reshuffled order, and multiple jobs often read shared datasets concurrently. The \texttt{md5\_check} benchmark reproduces these patterns with configurable sequential, seeded-random, and strided access. Checksumming forces every byte through the full read path so the measurement reflects storage performance rather than skipped I/O, and page-cache bypass via \texttt{posix\_fadvise(DONTNEED)} isolates true storage behavior from operating-system caching effects.

\subsubsection{Model Checkpointing}

Checkpointing persists training state for fault recovery and produces the primary artifact of a training job, exercising the large-I/O regime. \sysname{} covers the two dominant strategies: centralized checkpointing for data-parallel training, where a single rank writes the full model state to one file and all ranks read it back (\texttt{ddp\_save}, \texttt{ddp\_load}), and distributed checkpointing for fully sharded data-parallel training, where each rank writes its own shard and restore requires coordinated reads of all shards (\texttt{fsdp\_save}, \texttt{fsdp\_load}). Both support a configurable synthetic model for controlled size sweeps and a range of Hugging Face architectures (e.g., GPT-2, Llama, BERT variants) for realistic validation. 
The \texttt{rl\_load} benchmark covers the asymmetric reinforcement-learning pattern in which many training ranks hold sharded checkpoints while rollout workers need the consolidated full model, exercising the shard-consolidation read path.

\subsubsection{Dataset Generation}

Researchers routinely generate synthetic datasets, from simulation rollouts and trajectories to model-synthesized corpora and procedurally augmented data. Generation is write-dominated: large volumes of write-once files across a wide size range, produced as fast as the compute pipeline can emit them. The \texttt{create\_dataset} benchmark models the storage-facing side of this workflow, writing files from 8KB to 1GB to span the small- and large-I/O regimes identified in our characterization, with multiple files per size class to reach a target total dataset size. File content is generated deterministically via SHA256 hashing from a seed, keeping datasets reproducible across runs for verification and comparison.

\begin{table*}[t]
\centering
\resizebox{\linewidth}{!}{%
\begin{tabular}{r|l|c|l|l}
\rowcolor[HTML]{EFEFEF}
\textbf{Benchmark} & \multicolumn{1}{c|}{\cellcolor[HTML]{EFEFEF}\textbf{Description}} & \textbf{Category} & \multicolumn{1}{c|}{\cellcolor[HTML]{EFEFEF}\textbf{Access Pattern}} & \multicolumn{1}{|c}{\cellcolor[HTML]{EFEFEF}\textbf{Key Configuration}} \\ \hline

\texttt{ddp\_save} & Train iteration followed by rank 0 checkpoint save & Checkpointing & Single rank writes & Synthetic or HuggingFace model \\
\texttt{ddp\_load} & All ranks load same checkpoint, run training iteration & Checkpointing & All ranks read same file & Verifies via parameter checksum \\
\texttt{fsdp\_save} & All ranks write shards in parallel via DCP & Checkpointing & All ranks write different files & Creates sharded checkpoint directory \\
\texttt{fsdp\_load} & All ranks load individual shards via DCP & Checkpointing & All ranks read different files & Supports full state dict option \\
\texttt{rl\_load} & FSDP load with state dict consolidation & Checkpointing & All ranks read and consolidate & For RL inference workers \\ \hline

\texttt{md5\_check} & Read files and compute MD5 checksums & Dataloading & Sequential, random, or strided & Chunk size, page cache control \\
\texttt{create\_dataset} & Generate synthetic files with deterministic content & Data generation & Single rank writes & File sizes from 8KB to 1GB \\
\texttt{create\_synthetic\_workload} & Generate files with configurable distributions & Data generation & Single rank writes & Arrival, size, data patterns \\ \hline
\texttt{create\_files} & Create N files of specified size & Metadata & Per-rank or shared directory & File size, page cache bypass \\
\texttt{list\_files} & Create files then stat each file & Metadata & Per-rank directories & Measures readdir and stat \\
\texttt{move\_files} & Create then move N files between directories & Metadata & Per-rank directories & Measures rename latency \\
\texttt{delete\_files} & Create then delete N files & Metadata & Per-rank directories & Measures unlink latency \\
\texttt{folder\_bench} & Create nested directory hierarchies & Metadata & Per-rank directories & Depth and breadth configurable \\ \hline

\texttt{git\_clone} & Clone repository N times & Mixed & Per-rank directories & Repository URL configurable \\

\end{tabular}
}
\caption{Complete benchmark taxonomy in \sysname{}, organized by category with access patterns and configuration options.}
\label{tab:fsbench-benchmarks}
\end{table*}

\subsection{Framework Architecture}

\sysname{} employs a three-layer architecture: (i) a command-line interface that parses common and module-specific arguments, (ii) a core measurement layer that provides timing infrastructure through Python decorators, and (iii) a set of pluggable benchmark modules. The interface uses a two-stage parser: common arguments such as dataset path and output directory are parsed first, followed by module-specific arguments, so each module defines its own parameters without modifying the core.

The measurement layer exposes two decorators via the \texttt{BenchParams} class. \texttt{@benchable} registers functions in a global registry for discovery and dispatch by name; at invocation the framework introspects the function signature and binds command-line arguments to parameters. \texttt{@measurable} wraps functions with high-resolution timing using nanosecond-precision counters, accumulating measurements across invocations for statistical analysis. The decorators compose: an outer \texttt{@benchable} function defines the benchmark entry point, while inner \texttt{@measurable} functions capture fine-grained sub-operation timings (see listing below).

\begin{lstlisting}[caption={The benchable decorator registers functions for dispatch by name; the measurable decorator captures sub-operation timing with nanosecond precision.}]
@BenchParams.benchable
def create_files(num_copies: int,
                 dataset_path: str,
                 file_size: int = 4096) -> bool:
    for i in range(num_copies):
        _create_and_write_file(dataset_path, i, file_size)
    return True

@BenchParams.measurable
def _create_and_write_file(path: str, idx: int, size: int):
    filename = os.path.join(path, f"file_{idx}.dat")
    with open(filename, "wb") as f:
        f.write(b"\xab" * size)
\end{lstlisting}

After execution, the framework computes mean, minimum, maximum, and percentiles (p50, p90, p99) for each measured operation. Because research workloads are interactive and bursty, \sysname{} emphasizes latency percentiles over peak throughput: tail latency determines user-perceived performance. Results are serialized to JSON manifests recording input parameters, host metadata, timing statistics, and pass/fail status, where failure covers exceptions, timeouts, and output validation mismatches.

\subsection{Distributed Execution}

Distributed benchmarks coordinate multiple processes, typically one per GPU. \sysname{} relies on cluster schedulers such as SLURM for gang scheduling so that all ranks execute synchronously.

Process-group initialization supports two methods. The environment-variable method uses scheduler-provided master address and port, suitable for most cluster deployments. The file-based method coordinates through a shared filesystem path, useful when environment variables are unavailable.

For distributed checkpointing, \sysname{} follows a barrier-based coordination pattern: ranks synchronize before training to confirm readiness, after training to confirm completion, and after checkpoint I/O to confirm all writes finished before proceeding. This isolates storage performance from process-scheduling variation. The framework selects GPUs by local rank, chooses between the NCCL backend for GPU clusters and Gloo for CPU-only environments, and derives global rank and world size from scheduler environment variables. Each rank writes its own JSON result file, so post hoc aggregation can separate per-rank variation from aggregate system behavior.

\subsection{Extending \sysname{} with New Benchmarks}

Each benchmark module implements a plugin interface: a list of available functions, a module-specific argument parser, and an execution entry point. New benchmarks are added by implementing this interface and registering the module in the module list. \sysname{} integrates existing tools such as fio through the same interface for peak I/O measurements, spanning synthetic microbenchmarks to AI-specific workloads.

The listing below shows the extension pattern for loading a Hugging Face model with distributed coordination: rank 0 populates the shared cache, a single barrier synchronizes all ranks, and every rank then loads from the cache.

\begin{lstlisting}[caption={Extension example: Hugging Face model loading with distributed coordination.}]
@BenchParams.benchable
def hf_model_load(model_name: str,
                  cache_dir: str,
                  batch_per_rank: int) -> bool:
    # Rank 0 populates the shared cache
    if dist.is_initialized() and dist.get_rank() == 0:
        AutoModelForCausalLM.from_pretrained(
            model_name, cache_dir=cache_dir)

    # All ranks synchronize, then load from the shared cache
    if dist.is_initialized():
        dist.barrier()
    model = AutoModelForCausalLM.from_pretrained(
        model_name, cache_dir=cache_dir)

    return setup_training(model, batch_per_rank)
\end{lstlisting}



\newcommand{\LustreSystem}{\texttt{Lustre\_SYS}}  
\newcommand{\NASSystem}{\texttt{NAS\_SYS}}         
\newcommand{\ClusterName}{\texttt{CLUSTER\_NAME}}
\newcommand{\VendorName}{\texttt{VENDOR\_NAME}}

\section{Using \sysname{}}
\label{sec:using}

We built \sysname{} so infrastructure teams can systematically validate and qualify POSIX storage for AI research clusters across job sizes, benchmarks, and cluster configurations. We demonstrate four use cases from operating large-scale AI research storage: full-suite characterization of a single system (\S~\ref{sec:perf_char}); Lustre versus NAS appliance across workload profiles (\S~\ref{sec:lustre_vs_nas}); detecting a regression from a storage upgrade (\S~\ref{sec:regression}); and \texttt{fsspec} versus native POSIX access (\S~\ref{sec:fsspec}).

\subsection{Performance and Scalability Characterization}
\label{sec:perf_char}

We ran the full suite on \texttt{8K} Nvidia Hopper class GPUs backed by a \NASSystem\ appliance (\texttt{8}~PB; \texttt{150}~GB/s peak sequential bandwidth; over \texttt{1M} IOPS) to baseline performance and qualify the system for research workloads. Checkpointing (\texttt{ddp\_save}, \texttt{ddp\_load}, \texttt{fsdp\_save}, \texttt{fsdp\_load}), data loading (\texttt{md5\_check}), and metadata (\texttt{create\_files}, \texttt{list\_files}, \texttt{move\_files}, \texttt{delete\_files}) benchmarks ran at \texttt{8}--\texttt{256} GPUs. The same methodology guides storage selection and readiness qualification during cluster bring-ups.


%

\subsubsection{Checkpoint Performance}

Checkpoint save/restore latency directly affects cluster GPU utilization, so we evaluate it first. Figure~\ref{fig:lustre_nas_ckpt} presents the results for both DDP and FSDP checkpointing strategies across increasing task sizes.

\texttt{ddp\_save} latency remains stable from \texttt{8} to \texttt{256} GPUs, with mean latency of 7700--8000~ms at the largest scale. \texttt{ddp\_load} holds at 4000~ms mean latency up to \texttt{128} GPUs, but P99 latency rises to 10000~ms at \texttt{256} GPUs, indicating growing performance variance under concurrent load at scale. The \texttt{fsdp\_save} and \texttt{fsdp\_load} benchmarks, which spread I/O across all ranks, show gradually decreasing latency with scale: P99 (straggler-rank) latency falls from 1600~ms to 500~ms for save and from 2500~ms to 600~ms for load. These results confirm that the storage system sustains the I/O demands of distributed checkpointing at the target cluster scale, especially for highly parallel checkpointing styles.

\subsubsection{Data-Loading Performance}

We evaluate data-loading performance with the \texttt{md5\_check} benchmark, which reads files and computes checksums, exercising the sequential and random read patterns representative of training data access. We compare four configurations: (i)~\texttt{nas}, the vanilla variant using the page cache with sequential 8~KB-chunk access; (ii)~\texttt{nas\_nocache}, bypassing the page cache; (iii)~\texttt{nas\_nocache\_random}, uncached with random access order; and (iv)~\texttt{nas\_nocache\_random\_32K}, random order with 32~KB chunks.

\begin{figure}[ht]
\centering
\begin{tikzpicture}
\begin{axis}[
    ybar=1pt,
    bar width=4pt,
    width=0.95\columnwidth,
    height=0.45\columnwidth,
    enlarge x limits=0.12,
    symbolic x coords={8,16,32,64,128,256},
    xtick=data,
    label style={font=\footnotesize},
    tick label style={font=\footnotesize},
    legend style={
        at={(0,-.2)},
        anchor=north west,
        font=\scriptsize,
        draw=none,
        legend columns=2,
        column sep=1em,
    },
    legend cell align=left,
    area legend,
    ymin=0,
    ymax=290000,
    ytick={0,50000,100000,150000,200000,250000},
    yticklabels={0,50k,100k,150k,200k,250k},
    scaled y ticks=false,
    y tick label style={/pgf/number format/fixed, /pgf/number format/1000 sep={}},
    ymajorgrids=true,
    grid style={dashed, gray!40},
]

\addplot[fill=tabblue, draw=black,
    error bars/y dir=both, error bars/y explicit]
    coordinates {
        (8, 230000) +- (0, 3000)
        (16, 228000) +- (0, 3000)
        (32, 232000) +- (0, 3000)
        (64, 230000) +- (0, 3000)
        (128, 231000) +- (0, 3000)
        (256, 232000) +- (0, 5000)
    };
\addlegendentry{\texttt{nas}}

\addplot[fill=taborange, draw=black,
    error bars/y dir=both, error bars/y explicit]
    coordinates {
        (8, 261000) +- (0, 5000)
        (16, 239000) +- (0, 4000)
        (32, 240000) +- (0, 5000)
        (64, 244000) +- (0, 5000)
        (128, 239000) +- (0, 5000)
        (256, 240000) +- (0, 8000)
    };
\addlegendentry{\texttt{nas\_nocache}}

\addplot[fill=tabgreen, draw=black,
    error bars/y dir=both, error bars/y explicit]
    coordinates {
        (8, 260000) +- (0, 5000)
        (16, 253000) +- (0, 5000)
        (32, 268000) +- (0, 5000)
        (64, 242000) +- (0, 5000)
        (128, 240000) +- (0, 5000)
        (256, 241000) +- (0, 8000)
    };
\addlegendentry{\texttt{nas\_nocache\_random}}

\addplot[fill=tabred, draw=black,
    error bars/y dir=both, error bars/y explicit]
    coordinates {
        (8, 260000) +- (0, 5000)
        (16, 244000) +- (0, 5000)
        (32, 265000) +- (0, 5000)
        (64, 240000) +- (0, 5000)
        (128, 240000) +- (0, 5000)
        (256, 243000) +- (0, 10000)
    };
\addlegendentry{\texttt{nas\_nocache\_random\_32K}}

\end{axis}
\end{tikzpicture}%
\caption{Dataloading latency characterization using the \texttt{md5\_check} benchmark across increasing job scales under different settings. The $x$-axis shows the task count and the $y$-axis shows latency in milliseconds (ms). Error bars represent 95\% confidence intervals.}
\label{fig:dataload_perf}
\end{figure}

Figure~\ref{fig:dataload_perf} shows data-loading behavior across task sizes under each configuration. Trends are consistent across task sizes, indicating that the \NASSystem{} sustains this data-loading load with ease. Uncached latency (\texttt{nas\_nocache}) is 4--13\% higher than cached latency (\texttt{nas}), showing the OS page cache helps data loading. Random access adds a 9--13\% penalty over cached sequential latency, and the larger chunk size adds a further 6--15\%.

\subsubsection{File Metadata Performance}

Figure~\ref{fig:lustre_nas_meta} presents the latency of the four metadata benchmarks across increasing client counts.

As shown in Figure~\ref{fig:lustre_nas_meta}, \texttt{create\_files} and \texttt{move\_files} latency increases by 4.8~$\times$ and 12~$\times$ respectively as the client count grows from \texttt{8} to \texttt{256}. The \texttt{list\_files} operation is the least sensitive to scale, with latency unchanged as the client count grows. These results establish the metadata performance envelope and identify bottlenecks for metadata-heavy workflows such as environment setup and data preparation.

\subsubsection{Checkpoint Loading for RL}
\label{sec:rl_load}

Reinforcement-learning pipelines exercise an asymmetric checkpoint path: training ranks write sharded state, but rollout workers must read back a consolidated full model. The \texttt{rl\_load} benchmark reproduces this shard-consolidation read, and Figure~\ref{fig:rl_load} reports its latency and logical throughput on both systems.

\begin{figure}[ht]
    \centering
    \begin{subfigure}{0.49\columnwidth}
        \begin{tikzpicture}
\begin{axis}[
    ybar=0.5pt,
    bar width=1.8pt,
    width=\textwidth,
    height=0.85\textwidth,
    enlarge x limits=0.11,
    symbolic x coords={8,16,32,64,128,256},
    xtick=data,
    label style={font=\footnotesize},
    tick label style={font=\footnotesize},
    x tick label style={font=\tiny},
    ymin=0,
    ymax=60,
    ytick={0,20,40,60},
    scaled y ticks=false,
    ymajorgrids=true,
    grid style={dashed, gray!40},
]

\addplot[fill=tabblue, draw=black]
    coordinates {(8,6.9)(16,9.1)(32,11.2)(64,11.2)(128,15.2)(256,19.8)};

\addplot[fill=tabbluelight, draw=black,
    postaction={pattern=horizontal lines, pattern color=tabblue!50}]
    coordinates {(8,7.1)(16,12.4)(32,12.2)(64,11.2)(128,17.9)(256,21.5)};

\addplot[fill=taborange, draw=black]
    coordinates {(8,14.6)(16,7.4)(32,9.5)(64,15.2)(128,14.7)(256,20.6)};

\addplot[fill=taborangelight, draw=black,
    postaction={pattern=horizontal lines, pattern color=taborange!50}]
    coordinates {(8,34.1)(16,13.9)(32,14.6)(64,52.5)(128,33.5)(256,28.7)};

\end{axis}
\end{tikzpicture}%
        \caption{Checkpoint load latency}
    \end{subfigure}
    \begin{subfigure}{0.49\columnwidth}
        \begin{tikzpicture}
\begin{axis}[
    ybar=1pt,
    bar width=3.5pt,
    width=\textwidth,
    height=0.85\textwidth,
    enlarge x limits=0.11,
    symbolic x coords={8,16,32,64,128,256},
    xtick=data,
    label style={font=\footnotesize},
    tick label style={font=\footnotesize},
    x tick label style={font=\tiny},
    ymin=0,
    ymax=0.8,
    ytick={0,0.2,0.4,0.6,0.8},
    scaled y ticks=false,
    ymajorgrids=true,
    grid style={dashed, gray!40},
]

\addplot[fill=tabblue, draw=black]
    coordinates {(8,0.61)(16,0.47)(32,0.38)(64,0.38)(128,0.28)(256,0.22)};

\addplot[fill=taborange, draw=black]
    coordinates {(8,0.29)(16,0.58)(32,0.45)(64,0.28)(128,0.29)(256,0.21)};

\end{axis}
\end{tikzpicture}%
        \caption{Logical throughput}
    \end{subfigure}
    \\[0.3em]
    \begin{tikzpicture}
    \draw[fill=tabblue, draw=black] (0,0.3) rectangle (0.3,0.5);
    \node[right, font=\small] at (0.35,0.4) {\texttt{LUSTRE\_SYS} (Mean)};
    \draw[fill=tabbluelight, draw=black, postaction={pattern=horizontal lines, pattern color=tabblue!50}] (4.1,0.3) rectangle (4.4,0.5);
    \node[right, font=\small] at (4.45,0.4) {\texttt{LUSTRE\_SYS} (P99)};
    \draw[fill=taborange, draw=black] (0,0) rectangle (0.3,0.2);
    \node[right, font=\small] at (0.35,0.1) {\texttt{NAS\_SYS} (Mean)};
    \draw[fill=taborangelight, draw=black, postaction={pattern=horizontal lines, pattern color=taborange!50}] (4.1,0) rectangle (4.4,0.2);
    \node[right, font=\small] at (4.45,0.1) {\texttt{NAS\_SYS} (P99)};
\end{tikzpicture}
    \caption{
    Shard-consolidation checkpoint loading (\texttt{rl\_load}) on \LustreSystem{} and \NASSystem{} using \sysname{}, across increasing task counts.
    The $x$-axis shows the task count in both subgraphs; the $y$-axis shows (a)~load latency in seconds (s) and (b)~aggregate logical throughput in GB/s.
    Only mean values are reported in (b).
    }
    \label{fig:rl_load}
\end{figure}

Mean load latency grows modestly with scale on both systems---from 6.9~s to 19.8~s on \LustreSystem{} and from 14.6~s to 20.6~s on \NASSystem{} between \texttt{8} and \texttt{256} tasks---but the tails behave very differently. \NASSystem{} P99 latency reaches 52.5~s at \texttt{64} tasks and 34.1~s at \texttt{8} tasks, 2--3~$\times$ its own mean, whereas \LustreSystem{} P99 tracks within 30\% of its mean at every task count. Logical throughput correspondingly falls from 0.61~GB/s to 0.22~GB/s on \LustreSystem{} and from 0.29~GB/s to 0.21~GB/s on \NASSystem{} as consolidation work is spread over more readers. Shard consolidation is therefore straggler-dominated rather than bandwidth-dominated, and the P99 is the metric that governs how quickly rollout workers can resume.

\subsection{Comparing Storage Architectures: Lustre vs.\ NAS}
\label{sec:lustre_vs_nas}

Our clusters deploy two POSIX storage architectures: \LustreSystem\, a Lustre-based parallel filesystem, and \NASSystem\, a NAS appliance. Lustre targets high-throughput parallel I/O with a distributed metadata architecture, while the NAS appliance trades POSIX strictness for simpler deployment; their scaling characteristics differ accordingly. We ran the full \sysname{} suite on both systems in comparable cluster configurations, scaling from \texttt{8} to \texttt{256} GPUs. Figure~\ref{fig:lustre_nas_ckpt} presents the checkpointing results, and Figure~\ref{fig:lustre_nas_meta} presents the metadata results.

\subsubsection{Checkpoint operations}

\begin{figure*}[h]
    \centering
    \begin{subfigure}{0.24\textwidth}
        \begin{tikzpicture}
\begin{axis}[
    ybar=0.5pt,
    bar width=1.6pt,
    width=\textwidth,
    height=0.85\textwidth,
    enlarge x limits=0.11,
    symbolic x coords={8,16,32,64,128,256},
    xtick=data,
    label style={font=\footnotesize},
    tick label style={font=\footnotesize},
    x tick label style={font=\tiny},
    ymin=0,
    ymax=12000,
    scaled y ticks=false,
    y tick label style={/pgf/number format/fixed, /pgf/number format/1000 sep={}},
    ymajorgrids=true,
    grid style={dashed, gray!40},
]
\addplot[fill=tabblue, draw=black] coordinates {(8,5250)(16,5200)(32,5250)(64,5250)(128,5250)(256,5600)};
\addplot[fill=tabbluelight, draw=black, postaction={pattern=horizontal lines, pattern color=tabblue!50}] coordinates {(8,5750)(16,5550)(32,5700)(64,5850)(128,5900)(256,6250)};
\addplot[fill=taborange, draw=black] coordinates {(8,4850)(16,4500)(32,4100)(64,3950)(128,4200)(256,7700)};
\addplot[fill=taborangelight, draw=black, postaction={pattern=horizontal lines, pattern color=taborange!50}] coordinates {(8,5650)(16,5300)(32,4450)(64,4300)(128,5100)(256,10500)};
\end{axis}
\end{tikzpicture}%
        \caption{DDP Load}
    \end{subfigure}
    \begin{subfigure}{0.24\textwidth}
        \begin{tikzpicture}
\begin{axis}[
    ybar=0.5pt,
    bar width=1.6pt,
    width=\textwidth,
    height=0.85\textwidth,
    enlarge x limits=0.11,
    symbolic x coords={8,16,32,64,128,256},
    xtick=data,
    label style={font=\footnotesize},
    tick label style={font=\footnotesize},
    x tick label style={font=\tiny},
    ymin=0,
    ymax=9000,
    scaled y ticks=false,
    y tick label style={/pgf/number format/fixed, /pgf/number format/1000 sep={}},
    ymajorgrids=true,
    grid style={dashed, gray!40},
]
\addplot[fill=tabblue, draw=black] coordinates {(8,5500)(16,5500)(32,5500)(64,5700)(128,5600)(256,6200)};
\addplot[fill=tabbluelight, draw=black, postaction={pattern=horizontal lines, pattern color=tabblue!50}] coordinates {(8,5600)(16,5850)(32,5950)(64,6100)(128,6350)(256,7150)};
\addplot[fill=taborange, draw=black] coordinates {(8,7100)(16,7350)(32,7200)(64,7150)(128,7500)(256,7600)};
\addplot[fill=taborangelight, draw=black, postaction={pattern=horizontal lines, pattern color=taborange!50}] coordinates {(8,7300)(16,7950)(32,7400)(64,8100)(128,7800)(256,7700)};
\end{axis}
\end{tikzpicture}%
        \caption{DDP Save}
    \end{subfigure}
    \begin{subfigure}{0.24\textwidth}
        \begin{tikzpicture}
\begin{axis}[
    ybar=0.5pt,
    bar width=1.6pt,
    width=\textwidth,
    height=0.85\textwidth,
    enlarge x limits=0.11,
    symbolic x coords={8,16,32,64,128,256},
    xtick=data,
    label style={font=\footnotesize},
    tick label style={font=\footnotesize},
    x tick label style={font=\tiny},
    ymin=0,
    ymax=4500,
    scaled y ticks=false,
    y tick label style={/pgf/number format/fixed, /pgf/number format/1000 sep={}},
    ymajorgrids=true,
    grid style={dashed, gray!40},
]
\addplot[fill=tabblue, draw=black] coordinates {(8,2250)(16,250)(32,870)(64,830)(128,600)(256,450)};
\addplot[fill=tabbluelight, draw=black, postaction={pattern=horizontal lines, pattern color=tabblue!50}] coordinates {(8,4100)(16,290)(32,1000)(64,1050)(128,1400)(256,650)};
\addplot[fill=taborange, draw=black] coordinates {(8,1270)(16,730)(32,400)(64,250)(128,260)(256,310)};
\addplot[fill=taborangelight, draw=black, postaction={pattern=horizontal lines, pattern color=taborange!50}] coordinates {(8,1650)(16,1430)(32,480)(64,350)(128,680)(256,520)};
\end{axis}
\end{tikzpicture}%
        \caption{FSDP Load}
    \end{subfigure}
    \begin{subfigure}{0.24\textwidth}
        \begin{tikzpicture}
\begin{axis}[
    ybar=0.5pt,
    bar width=1.6pt,
    width=\textwidth,
    height=0.85\textwidth,
    enlarge x limits=0.11,
    symbolic x coords={8,16,32,64,128,256},
    xtick=data,
    label style={font=\footnotesize},
    tick label style={font=\footnotesize},
    x tick label style={font=\tiny},
    ymin=0,
    ymax=2800,
    scaled y ticks=false,
    y tick label style={/pgf/number format/fixed, /pgf/number format/1000 sep={}},
    ymajorgrids=true,
    grid style={dashed, gray!40},
]
\addplot[fill=tabblue, draw=black] coordinates {(8,2000)(16,1100)(32,850)(64,830)(128,530)(256,560)};
\addplot[fill=tabbluelight, draw=black, postaction={pattern=horizontal lines, pattern color=tabblue!50}] coordinates {(8,2100)(16,1150)(32,980)(64,1030)(128,720)(256,680)};
\addplot[fill=taborange, draw=black] coordinates {(8,2220)(16,1270)(32,680)(64,400)(128,350)(256,500)};
\addplot[fill=taborangelight, draw=black, postaction={pattern=horizontal lines, pattern color=taborange!50}] coordinates {(8,2550)(16,1480)(32,850)(64,540)(128,510)(256,740)};
\end{axis}
\end{tikzpicture}%
        \caption{FSDP Save}
    \end{subfigure}
    \\[0.3em]
    \begin{tikzpicture}
    \draw[fill=tabblue, draw=black] (0,0) rectangle (0.3,0.2);
    \node[right, font=\small] at (0.35,0.1) {\texttt{LUSTRE\_SYS} (Mean)};
    \draw[fill=tabbluelight, draw=black, postaction={pattern=horizontal lines, pattern color=tabblue!50}] (3.8,0) rectangle (4.1,0.2);
    \node[right, font=\small] at (4.15,0.1) {\texttt{LUSTRE\_SYS} (P99)};
    \draw[fill=taborange, draw=black] (7.6,0) rectangle (7.9,0.2);
    \node[right, font=\small] at (7.95,0.1) {\texttt{NAS\_SYS} (Mean)};
    \draw[fill=taborangelight, draw=black, postaction={pattern=horizontal lines, pattern color=taborange!50}] (11,0) rectangle (11.3,0.2);
    \node[right, font=\small] at (11.35,0.1) {\texttt{NAS\_SYS} (P99)};
\end{tikzpicture}
    \caption{Checkpoint load/save performance characterization on \LustreSystem{} and \NASSystem{} using \sysname{}. $x$-axis: task count, $y$-axis: checkpoint latency in milliseconds. Error bars: 95\% CI.}
    \label{fig:lustre_nas_ckpt}
\end{figure*}

The \NASSystem{} outperforms \LustreSystem{} on checkpoint loads, while \LustreSystem{} is competitive on saves thanks to striping writes across object storage targets. For \texttt{ddp\_save}, \LustreSystem{} outperforms \NASSystem{} by 9--35\%. 
For \texttt{fsdp\_save}, \LustreSystem{} is up to 26\% faster at small task counts but up to 2~$\times$ slower at large ones. On loads, \NASSystem{} leads across most task counts: up to 20\% for \texttt{ddp\_load} and 3~$\times$ for \texttt{fsdp\_load}.

\subsubsection{Data-Loading Operations}

We compare data-loading time via the \texttt{md5\_check} benchmark with random file access over a synthetic dataset of 800K 32~KiB files in a single flat directory. \LustreSystem{} handles many files per directory poorly due to limited metadata performance; the standard remedy is sharding files across directories to disperse metadata overhead.

\begin{figure}[h]
\centering
\begin{tikzpicture}
\begin{axis}[
    width=0.8\columnwidth,
    height=0.5\columnwidth,
    xmin=-1.0, xmax=1.0,
    xtick={-0.6,0,0.6},
    xticklabels={\NASSystem, \LustreSystem, {\LustreSystem\ (sharded)}},
    x tick label style={font=\tiny},
    tick label style={font=\footnotesize},
    ymin=0,
    ymax=22000,
    ytick={0,5000,10000,15000,20000},
    scaled y ticks=false,
    y tick label style={/pgf/number format/fixed, /pgf/number format/1000 sep={}},
    ymajorgrids=true,
    grid style={dashed, gray!40},
]

\addplot[ybar, bar width=20pt, fill=tabblue, draw=black,
    error bars/y dir=both, error bars/y explicit]
    coordinates {(-0.75, 3190) +- (0, 60)};

\addplot[ybar, bar width=20pt, fill=tabbluelight, draw=black,
    postaction={pattern=horizontal lines, pattern color=tabblue!50}]
    coordinates {(-0.45, 3190)};

\addplot[ybar, bar width=20pt, fill=taborange, draw=black,
    error bars/y dir=both, error bars/y explicit]
    coordinates {(-0.15, 16910) +- (0, 4930)};

\addplot[ybar, bar width=20pt, fill=taborangelight, draw=black,
    postaction={pattern=horizontal lines, pattern color=taborange!50}]
    coordinates {(0.15, 16910)};

\addplot[ybar, bar width=20pt, fill=tabgreen, draw=black,
    error bars/y dir=both, error bars/y explicit]
    coordinates {(0.45, 5950) +- (0, 60)};

\addplot[ybar, bar width=20pt, fill=tabgreenlight, draw=black,
    postaction={pattern=horizontal lines, pattern color=tabgreen!50}]
    coordinates {(0.75, 5950)};

\end{axis}
\end{tikzpicture}%
\caption{Dataloading performance (tasks$=$\texttt{256}) showing latency in seconds, synthetic dataset of 800K small files (32KiB). \NASSystem{} (blue) vs. \LustreSystem{} wo/ sharding (orange) and w/ sharding (green). Solid bar: Mean, Light bar: P99, Error bars: 95\% CI.
Sharding datasets for \LustreSystem{} helps mitigate the metadata performance bottleneck.}
\label{fig:dataloading_lustre_vs_nas}
\end{figure}

As shown in Figure~\ref{fig:dataloading_lustre_vs_nas}, \LustreSystem{} is over 5~$\times$ slower than \NASSystem{} at \texttt{256} tasks. Sharding the dataset across directories narrows the gap, improving \LustreSystem{} performance by nearly 3~$\times$ over the unsharded layout.

\subsubsection{Metadata Operations}

\begin{figure*}[h]
\centering
    \begin{subfigure}{0.24\textwidth}
        \begin{tikzpicture}
\begin{axis}[
    ybar=0.5pt,
    bar width=1.6pt,
    width=\textwidth,
    height=0.85\textwidth,
    enlarge x limits=0.11,
    symbolic x coords={8,16,32,64,128,256},
    xtick=data,
    label style={font=\footnotesize},
    tick label style={font=\footnotesize},
    x tick label style={font=\tiny},
    ymin=0,
    ymax=580000,
    ytick={0,200000,400000},
    yticklabels={0,200k,400k},
    scaled y ticks=false,
    y tick label style={/pgf/number format/fixed, /pgf/number format/1000 sep={}},
    ymajorgrids=true,
    grid style={dashed, gray!40},
]
\addplot[fill=tabblue, draw=black] coordinates {(8,60000)(16,68000)(32,95000)(64,120000)(128,122000)(256,172000)};
\addplot[fill=tabbluelight, draw=black, postaction={pattern=horizontal lines, pattern color=tabblue!50}] coordinates {(8,62000)(16,70000)(32,97000)(64,122000)(128,124000)(256,175000)};
\addplot[fill=taborange, draw=black] coordinates {(8,112000)(16,110000)(32,108000)(64,122000)(128,270000)(256,520000)};
\addplot[fill=taborangelight, draw=black, postaction={pattern=horizontal lines, pattern color=taborange!50}] coordinates {(8,114000)(16,112000)(32,110000)(64,124000)(128,272000)(256,540000)};
\end{axis}
\end{tikzpicture}%
        \caption{\texttt{create\_files}}
    \end{subfigure}
    \begin{subfigure}{0.24\textwidth}
        \begin{tikzpicture}
\begin{axis}[
    ybar=0.5pt,
    bar width=1.6pt,
    width=\textwidth,
    height=0.85\textwidth,
    enlarge x limits=0.11,
    symbolic x coords={8,16,32,64,128,256},
    xtick=data,
    label style={font=\footnotesize},
    tick label style={font=\footnotesize},
    x tick label style={font=\tiny},
    ymin=0,
    ymax=9500,
    ytick={0,2000,4000,6000,8000},
    yticklabels={0,2k,4k,6k,8k},
    scaled y ticks=false,
    y tick label style={/pgf/number format/fixed, /pgf/number format/1000 sep={}},
    ymajorgrids=true,
    grid style={dashed, gray!40},
]
\addplot[fill=tabblue, draw=black] coordinates {(8,6700)(16,6200)(32,7050)(64,7350)(128,7350)(256,8300)};
\addplot[fill=tabbluelight, draw=black, postaction={pattern=horizontal lines, pattern color=tabblue!50}] coordinates {(8,6800)(16,6300)(32,7150)(64,7400)(128,7450)(256,8500)};
\addplot[fill=taborange, draw=black] coordinates {(8,140)(16,80)(32,90)(64,80)(128,100)(256,80)};
\addplot[fill=taborangelight, draw=black, postaction={pattern=horizontal lines, pattern color=taborange!50}] coordinates {(8,180)(16,100)(32,110)(64,100)(128,120)(256,100)};
\end{axis}
\end{tikzpicture}%
        \caption{\texttt{list\_files}}
    \end{subfigure}
    \begin{subfigure}{0.24\textwidth}
        \begin{tikzpicture}
\begin{axis}[
    ybar=0.5pt,
    bar width=1.6pt,
    width=\textwidth,
    height=0.85\textwidth,
    enlarge x limits=0.11,
    symbolic x coords={8,16,32,64,128,256},
    xtick=data,
    label style={font=\footnotesize},
    tick label style={font=\footnotesize},
    x tick label style={font=\tiny},
    ymin=0,
    ymax=580000,
    ytick={0,200000,400000},
    yticklabels={0,200k,400k},
    scaled y ticks=false,
    y tick label style={/pgf/number format/fixed, /pgf/number format/1000 sep={}},
    ymajorgrids=true,
    grid style={dashed, gray!40},
]
\addplot[fill=tabblue, draw=black] coordinates {(8,53000)(16,47000)(32,40000)(64,45000)(128,57000)(256,62000)};
\addplot[fill=tabbluelight, draw=black, postaction={pattern=horizontal lines, pattern color=tabblue!50}] coordinates {(8,56000)(16,50000)(32,43000)(64,48000)(128,59000)(256,65000)};
\addplot[fill=taborange, draw=black] coordinates {(8,35000)(16,38000)(32,63000)(64,133000)(128,305000)(256,530000)};
\addplot[fill=taborangelight, draw=black, postaction={pattern=horizontal lines, pattern color=taborange!50}] coordinates {(8,37000)(16,40000)(32,65000)(64,135000)(128,308000)(256,545000)};
\end{axis}
\end{tikzpicture}%
        \caption{\texttt{delete\_files}}
    \end{subfigure}
    \begin{subfigure}{0.24\textwidth}
        \begin{tikzpicture}
\begin{axis}[
    ybar=0.5pt,
    bar width=1.6pt,
    width=\textwidth,
    height=0.85\textwidth,
    enlarge x limits=0.13,
    symbolic x coords={8,16,32,64,128},
    xtick=data,
    label style={font=\footnotesize},
    tick label style={font=\footnotesize},
    x tick label style={font=\tiny},
    ymin=0,
    ymax=1950000,
    ytick={0,500000,1000000,1500000},
    yticklabels={0,500k,1M,1.5M},
    scaled y ticks=false,
    y tick label style={/pgf/number format/fixed, /pgf/number format/1000 sep={}},
    ymajorgrids=true,
    grid style={dashed, gray!40},
]
\addplot[fill=tabblue, draw=black] coordinates {(8,630000)(16,790000)(32,1770000)(64,585000)(128,620000)};
\addplot[fill=tabbluelight, draw=black, postaction={pattern=horizontal lines, pattern color=tabblue!50}] coordinates {(8,640000)(16,800000)(32,1800000)(64,590000)(128,630000)};
\addplot[fill=taborange, draw=black] coordinates {(8,50000)(16,60000)(32,130000)(64,270000)(128,620000)};
\addplot[fill=taborangelight, draw=black, postaction={pattern=horizontal lines, pattern color=taborange!50}] coordinates {(8,55000)(16,65000)(32,135000)(64,280000)(128,640000)};
\end{axis}
\end{tikzpicture}%
        \caption{\texttt{move\_files}}
    \end{subfigure}
    \\[0.3em]
    \begin{tikzpicture}
    \draw[fill=tabblue, draw=black] (0,0) rectangle (0.3,0.2);
    \node[right, font=\small] at (0.35,0.1) {\texttt{LUSTRE\_SYS} (Mean)};
    \draw[fill=tabbluelight, draw=black, postaction={pattern=horizontal lines, pattern color=tabblue!50}] (3.8,0) rectangle (4.1,0.2);
    \node[right, font=\small] at (4.15,0.1) {\texttt{LUSTRE\_SYS} (P99)};
    \draw[fill=taborange, draw=black] (7.6,0) rectangle (7.9,0.2);
    \node[right, font=\small] at (7.95,0.1) {\texttt{NAS\_SYS} (Mean)};
    \draw[fill=taborangelight, draw=black, postaction={pattern=horizontal lines, pattern color=taborange!50}] (11,0) rectangle (11.3,0.2);
    \node[right, font=\small] at (11.35,0.1) {\texttt{NAS\_SYS} (P99)};
\end{tikzpicture}
\caption{Metadata performance comparison between \LustreSystem{} and \NASSystem{}. (a)~\texttt{create\_files}. (b)~\texttt{list\_files}. (c)~\texttt{delete\_files}. (d)~\texttt{move\_files}. 
$x$-axis: task count, $y$-axis: latency in milliseconds. Error bars: 95\% CI.}
\label{fig:lustre_nas_meta}
\end{figure*}

As shown in Figure~\ref{fig:lustre_nas_meta}, \NASSystem{} substantially outperforms \LustreSystem{} on metadata-intensive operations: \texttt{list\_files} is 80~$\times$ faster at \texttt{256} clients, nearly two orders of magnitude. Lustre's metadata scalability limitations~\cite{ior} grow more apparent at higher client counts.

No single storage architecture dominates across all workload categories: Lustre-based systems excel at large-file, throughput-intensive checkpointing, while NAS-based systems provide substantially better metadata performance for the interactive, small-file-heavy workflows of AI research. The divergence widens at scale, underscoring the need for workload-aware storage provisioning.


\subsection{Detecting Temporal Performance Regressions}
\label{sec:regression}

Production storage undergoes frequent upgrades---firmware patches, kernel updates, configuration changes, hardware replacements---and before-and-after benchmarks detect and quantify their unintended performance impact. We present a case study in which automated \sysname{} runs captured a severe checkpoint-latency regression after a routine vendor software update: the \texttt{fsdp\_load} benchmark regressed by approximately 4~$\times$ on small jobs (2 nodes) and up to 8~$\times$ on large jobs (128 nodes; not shown). Investigation traced the root cause to a bug in the vendor's NFS implementation: \texttt{openat} calls stalled for approximately 5 seconds when concurrent clients accessed the same file, and the resulting lock contention inflated checkpoint latency.

\begin{figure}[ht]
    \centering
    \begin{subfigure}{0.49\columnwidth}
        \begin{tikzpicture}
\begin{axis}[
    ybar=0.8pt,
    bar width=8pt,
    width=\textwidth,
    height=0.85\textwidth,
    enlarge x limits=0.35,
    symbolic x coords={8,16},
    xtick=data,
    label style={font=\footnotesize},
    tick label style={font=\footnotesize},
    ymin=0,
    ymax=7000,
    ytick={0,1000,2000,3000,4000,5000,6000},
    scaled y ticks=false,
    y tick label style={/pgf/number format/fixed, /pgf/number format/1000 sep={}},
    ymajorgrids=true,
    grid style={dashed, gray!40},
]

\addplot[fill=tabblue, draw=black,
    error bars/y dir=both, error bars/y explicit]
    coordinates {(8, 1150) +- (0, 50) (16, 891) +- (0, 60)};

\addplot[fill=tabbluelight, draw=black,
    postaction={pattern=horizontal lines, pattern color=tabblue!50}]
    coordinates {(8, 1380) (16, 1500)};

\addplot[fill=taborange, draw=black,
    error bars/y dir=both, error bars/y explicit]
    coordinates {(8, 1550) +- (0, 80) (16, 2109) +- (0, 200)};

\addplot[fill=taborangelight, draw=black,
    postaction={pattern=horizontal lines, pattern color=taborange!50}]
    coordinates {(8, 6400) (16, 6100)};

\end{axis}
\end{tikzpicture}%
        \caption{Checkpoint load (FSDP Load)}
    \end{subfigure}
    \begin{subfigure}{0.49\columnwidth}
        \begin{tikzpicture}
\begin{axis}[
    ybar=0.8pt,
    bar width=8pt,
    width=\textwidth,
    height=0.85\textwidth,
    enlarge x limits=0.35,
    symbolic x coords={8,16},
    xtick=data,
    label style={font=\footnotesize},
    tick label style={font=\footnotesize},
    ymin=0,
    ymax=5000,
    ytick={0,1000,2000,3000,4000},
    scaled y ticks=false,
    y tick label style={/pgf/number format/fixed, /pgf/number format/1000 sep={}},
    ymajorgrids=true,
    grid style={dashed, gray!40},
]

\addplot[fill=tabblue, draw=black,
    error bars/y dir=both, error bars/y explicit]
    coordinates {(8, 2600) +- (0, 60) (16, 2270) +- (0, 100)};

\addplot[fill=tabbluelight, draw=black,
    postaction={pattern=horizontal lines, pattern color=tabblue!50}]
    coordinates {(8, 3350) (16, 4500)};

\addplot[fill=taborange, draw=black,
    error bars/y dir=both, error bars/y explicit]
    coordinates {(8, 2620) +- (0, 70) (16, 2220) +- (0, 60)};

\addplot[fill=taborangelight, draw=black,
    postaction={pattern=horizontal lines, pattern color=taborange!50}]
    coordinates {(8, 3480) (16, 2850)};

\end{axis}
\end{tikzpicture}%
        \caption{Checkpoint save (FSDP Save)}
    \end{subfigure}
    \\[0.3em]
    \begin{tikzpicture}
    \draw[fill=tabblue, draw=black] (0,0.3) rectangle (0.3,0.5);
    \node[right, font=\small] at (0.35,0.4) {\texttt{NAS\_AFTER\_FIX} (Mean)};
    \draw[fill=tabbluelight, draw=black, postaction={pattern=horizontal lines, pattern color=tabblue!50}] (4.2,0.3) rectangle (4.5,0.5);
    \node[right, font=\small] at (4.55,0.4) {\texttt{NAS\_AFTER\_FIX} (P99)};
    \draw[fill=taborange, draw=black] (0,0) rectangle (0.3,0.2);
    \node[right, font=\small] at (0.35,0.1) {\texttt{NAS\_BEFORE\_FIX} (Mean)};
    \draw[fill=taborangelight, draw=black, postaction={pattern=horizontal lines, pattern color=taborange!50}] (4.2,0) rectangle (4.5,0.2);
    \node[right, font=\small] at (4.55,0.1) {\texttt{NAS\_BEFORE\_FIX} (P99)};
\end{tikzpicture}
    \caption{
    Performance regression detected by \sysname{} after a storage system upgrade on the \NASSystem{}.
    The latency before and after the fix are shown.
    In both subgraphs the $x$-axis shows the task count and the $y$-axis shows FSDP checkpoint latency in milliseconds (ms).
    Error bars represent 95\% confidence intervals.
    }
    \label{fig:regression}
\end{figure}

We reported the reproducible evidence to the vendor, who confirmed the bug and shipped a patch that \sysname{} validated before production rollout. Figure~\ref{fig:regression} confirms restoration to pre-upgrade levels: \texttt{fsdp\_load} mean latency at 16 tasks fell from 2109~ms to 891~ms, a 2.3~$\times$ improvement. Without continuous monitoring, the regression could have persisted for weeks, wasting GPU-hours on stalled checkpoint resumes.

The case study illustrates \sysname{}'s role in continuous integration: standardized benchmarks against established baselines, run after every system change, catch regressions early.

\subsection{Impact of the Storage Access Interface: \texttt{fsspec} vs.\ Native Access}
\label{sec:fsspec}

As discussed in \S~\ref{sec:fsspec_background}, \texttt{fsspec} provides a unified Python interface to diverse storage backends, simplifying application code at the cost of an extra software layer above the filesystem. 
We ran two configurations: (i)~Python native POSIX (e.g., \texttt{os.listdir()} and \texttt{os.stat()} for list\_files), and (ii)~the equivalent \texttt{fsspec} interface (e.g., \texttt{fs.ls()} and \texttt{fs.info()}) for a variety of common metadata heavy operations.
For e.g., we considered directory listing as it is frequent in data exploration, dataset validation, and experiment management, and its many individual metadata calls make it sensitive to per-operation overhead. 

\begin{figure*}[h]
\centering
    \begin{subfigure}{0.19\textwidth}
        \begin{tikzpicture}
\begin{axis}[
    width=\textwidth,
    height=1\textwidth,
    xmin=-1.2, xmax=1.2,
    xtick=\empty,
    tick label style={font=\scriptsize},
    scaled y ticks=false,
    ymin=0,
    ymax=160000,
    ytick={0,50000,100000,150000},
    yticklabels={0,50k,100k,150k},
    ymajorgrids=true,
    grid style={dashed, gray!40},
]

\addplot[ybar, bar width=8pt, fill=fsspecmean, draw=black,
    error bars/y dir=both, error bars/y explicit]
    coordinates {(-0.7, 140000) +- (0, 1200)};

\addplot[ybar, bar width=8pt, forget plot, fill=fsspecp99, draw=black,
    postaction={pattern=horizontal lines, pattern color=fsspecmean!50},
    error bars/y dir=both, error bars/y explicit]
    coordinates {(-0.25, 141000) +- (0, 1000)};

\addplot[ybar, bar width=8pt, fill=nativemean, draw=black,
    error bars/y dir=both, error bars/y explicit]
    coordinates {(0.25, 145000) +- (0, 1500)};

\addplot[ybar, bar width=8pt, forget plot, fill=nativep99, draw=black,
    postaction={pattern=horizontal lines, pattern color=nativemean!50},
    error bars/y dir=both, error bars/y explicit]
    coordinates {(0.7, 146000) +- (0, 1200)};

\end{axis}
\end{tikzpicture}%
        \caption{Create files}
    \end{subfigure}
    \begin{subfigure}{0.19\textwidth}
        \begin{tikzpicture}
\begin{axis}[
    width=\textwidth,
    height=1\textwidth,
    xmin=-1.2, xmax=1.2,
    xtick=\empty,
    tick label style={font=\scriptsize},
    scaled y ticks=false,
    ymin=0,
    ymax=230000,
    ytick={0,50000,100000,150000,200000},
    yticklabels={0,50k,100k,150k,200k},
    ymajorgrids=true,
    grid style={dashed, gray!40},
]

\addplot[ybar, bar width=8pt, fill=fsspecmean, draw=black,
    error bars/y dir=both, error bars/y explicit]
    coordinates {(-0.7, 205000) +- (0, 1500)};

\addplot[ybar, bar width=8pt, forget plot, fill=fsspecp99, draw=black,
    postaction={pattern=horizontal lines, pattern color=fsspecmean!50},
    error bars/y dir=both, error bars/y explicit]
    coordinates {(-0.25, 206000) +- (0, 1200)};

\addplot[ybar, bar width=8pt, fill=nativemean, draw=black,
    error bars/y dir=both, error bars/y explicit]
    coordinates {(0.25, 210000) +- (0, 2500)};

\addplot[ybar, bar width=8pt, forget plot, fill=nativep99, draw=black,
    postaction={pattern=horizontal lines, pattern color=nativemean!50},
    error bars/y dir=both, error bars/y explicit]
    coordinates {(0.7, 215000) +- (0, 2000)};

\end{axis}
\end{tikzpicture}%
        \caption{Move files}
    \end{subfigure}
    \begin{subfigure}{0.19\textwidth}
        \begin{tikzpicture}
\begin{axis}[
    width=\textwidth,
    height=1\textwidth,
    xmin=-1.2, xmax=1.2,
    xtick=\empty,
    tick label style={font=\scriptsize},
    scaled y ticks=false,
    ymin=0,
    ymax=140,
    ytick={0,40,80,120},
    yticklabels={0,40,80,120},
    ymajorgrids=true,
    grid style={dashed, gray!40},
]

\addplot[ybar, bar width=8pt, fill=fsspecmean, draw=black,
    error bars/y dir=both, error bars/y explicit]
    coordinates {(-0.7, 114) +- (0, 11.4)};

\addplot[ybar, bar width=8pt, forget plot, fill=fsspecp99, draw=black,
    postaction={pattern=horizontal lines, pattern color=fsspecmean!50},
    error bars/y dir=both, error bars/y explicit]
    coordinates {(-0.25, 114) +- (0, 11.4)};

\addplot[ybar, bar width=8pt, fill=nativemean, draw=black,
    error bars/y dir=both, error bars/y explicit]
    coordinates {(0.25, 103) +- (0, 8.9)};

\addplot[ybar, bar width=8pt, forget plot, fill=nativep99, draw=black,
    postaction={pattern=horizontal lines, pattern color=nativemean!50},
    error bars/y dir=both, error bars/y explicit]
    coordinates {(0.7, 103) +- (0, 8.9)};

\end{axis}
\end{tikzpicture}%
        \caption{List files}
    \end{subfigure}
    \begin{subfigure}{0.19\textwidth}
        \begin{tikzpicture}
\begin{axis}[
    width=\textwidth,
    height=1\textwidth,
    xmin=-1.2, xmax=1.2,
    xtick=\empty,
    tick label style={font=\scriptsize},
    scaled y ticks=false,
    ymin=0,
    ymax=200000,
    ytick={0,50000,100000,150000,200000},
    yticklabels={0,50k,100k,150k,200k},
    ymajorgrids=true,
    grid style={dashed, gray!40},
]

\addplot[ybar, bar width=8pt, fill=fsspecmean, draw=black,
    error bars/y dir=both, error bars/y explicit]
    coordinates {(-0.7, 157000) +- (0, 2000)};

\addplot[ybar, bar width=8pt, forget plot, fill=fsspecp99, draw=black,
    postaction={pattern=horizontal lines, pattern color=fsspecmean!50},
    error bars/y dir=both, error bars/y explicit]
    coordinates {(-0.25, 158000) +- (0, 1500)};

\addplot[ybar, bar width=8pt, fill=nativemean, draw=black,
    error bars/y dir=both, error bars/y explicit]
    coordinates {(0.25, 183000) +- (0, 2500)};

\addplot[ybar, bar width=8pt, forget plot, fill=nativep99, draw=black,
    postaction={pattern=horizontal lines, pattern color=nativemean!50},
    error bars/y dir=both, error bars/y explicit]
    coordinates {(0.7, 185000) +- (0, 2000)};

\end{axis}
\end{tikzpicture}%
        \caption{Delete files}
    \end{subfigure}
    \begin{subfigure}{0.19\textwidth}
        \begin{tikzpicture}
\begin{axis}[
    width=\textwidth,
    height=1\textwidth,
    xmin=-1.2, xmax=1.2,
    xtick=\empty,
    tick label style={font=\scriptsize},
    scaled y ticks=false,
    ymin=0,
    ymax=250000,
    ytick={0,50000,100000,150000,200000,250000},
    yticklabels={0,50k,100k,150k,200k,250k},
    ymajorgrids=true,
    grid style={dashed, gray!40},
]

\addplot[ybar, bar width=8pt, fill=fsspecmean, draw=black,
    error bars/y dir=both, error bars/y explicit]
    coordinates {(-0.7, 230000) +- (0, 2000)};

\addplot[ybar, bar width=8pt, forget plot, fill=fsspecp99, draw=black,
    postaction={pattern=horizontal lines, pattern color=fsspecmean!50},
    error bars/y dir=both, error bars/y explicit]
    coordinates {(-0.25, 231000) +- (0, 1500)};

\addplot[ybar, bar width=8pt, fill=nativemean, draw=black,
    error bars/y dir=both, error bars/y explicit]
    coordinates {(0.25, 220000) +- (0, 2500)};

\addplot[ybar, bar width=8pt, forget plot, fill=nativep99, draw=black,
    postaction={pattern=horizontal lines, pattern color=nativemean!50},
    error bars/y dir=both, error bars/y explicit]
    coordinates {(0.7, 222000) +- (0, 2000)};

\end{axis}
\end{tikzpicture}%
        \caption{MD5 check}
    \end{subfigure}
\begin{tikzpicture}
    \draw[fill=fsspecmean, draw=black] (0,0) rectangle (0.25,0.15);
    \node[right, font=\scriptsize] at (0.3,0.075) {\texttt{NAS\_FSSPEC} (Mean)};
    \draw[fill=fsspecp99, draw=black, postaction={pattern=horizontal lines, pattern color=fsspecmean!50}] (3,0) rectangle (3.25,0.15);
    \node[right, font=\scriptsize] at (3.3,0.075) {\texttt{NAS\_FSSPEC} (P99)};
    \draw[fill=nativemean, draw=black] (5.8,0) rectangle (6.05,0.15);
    \node[right, font=\scriptsize] at (6.1,0.075) {\texttt{NAS\_NATIVE} (Mean)};
    \draw[fill=nativep99, draw=black, postaction={pattern=horizontal lines, pattern color=nativemean!50}] (8.8,0) rectangle (9.05,0.15);
    \node[right, font=\scriptsize] at (9.1,0.075) {\texttt{NAS\_NATIVE} (P99)};
\end{tikzpicture}
\caption{Performance comparison of native POSIX access vs.\ \texttt{fsspec}, Task count $=$ 16. 
The \texttt{delete\_files} and \texttt{list\_files} benchmarks had non-negligible performance differences. 
$y$-axis: latency in milliseconds. Error bars: 95\% CI.}
\label{fig:fsspec}
\end{figure*}

Figure~\ref{fig:fsspec} shows only a modest gap for \texttt{list\_files}: for a directory of \texttt{5000} files, native access completes in 103~ms while \texttt{fsspec} requires 114~ms, an 11\% overhead attributable to \texttt{fsspec}'s abstraction layers, including filesystem-type parsing, dispatch, and result normalization into Python dictionaries. The 95\% confidence intervals of the two configurations overlap, so the difference is not statistically significant at this scale.

While higher level abstractions such as \texttt{fsspec} offer developer convenience and portability, it is additionally important to verify whether their performance is impacted by the storage backend.
More broadly, production slowdowns are often misattributed to the storage backend; \sysname{}'s interface optionality provides an A/B test of latency and throughput as observed by the workload, directing optimization effort at the right layer.

\subsection{Researcher Development Workflows}

\begin{figure*}[h]
    \centering
    \begin{subfigure}{0.24\textwidth}
        \begin{tikzpicture}
\begin{axis}[
    ybar=1pt,
    bar width=8pt,
    width=\textwidth,
    height=0.85\textwidth,
    enlarge x limits=0.35,
    symbolic x coords={1,4},
    xtick=data,
    label style={font=\footnotesize},
    tick label style={font=\footnotesize},
    ymin=0,
    ymax=31,
    ytick={0,10,20,30},
    scaled y ticks=false,
    ymajorgrids=true,
    grid style={dashed, gray!40},
]

\addplot[fill=tabblue, draw=black,
    error bars/y dir=both, error bars/y explicit]
    coordinates {
        (1, 12.83) += (0, 0.31) -= (0, 0.20)
        (4, 13.85) += (0, 0.24) -= (0, 0.12)
    };

\addplot[fill=tabbluelight, draw=black,
    postaction={pattern=horizontal lines, pattern color=tabblue!50}]
    coordinates {
        (1, 12.87)
        (4, 14.43)
    };

\addplot[fill=taborange, draw=black,
    error bars/y dir=both, error bars/y explicit]
    coordinates {
        (1, 22.91) += (0, 2.87) -= (0, 2.78)
        (4, 23.89) += (0, 4.65) -= (0, 4.69)
    };

\addplot[fill=taborangelight, draw=black,
    postaction={pattern=horizontal lines, pattern color=taborange!50}]
    coordinates {
        (1, 22.91)
        (4, 25.41)
    };

\end{axis}
\end{tikzpicture}%
        \caption{Git Clone}
    \end{subfigure}
    \begin{subfigure}{0.24\textwidth}
        \begin{tikzpicture}
\begin{axis}[
    ybar=1pt,
    bar width=8pt,
    width=\textwidth,
    height=0.85\textwidth,
    enlarge x limits=0.35,
    symbolic x coords={1,4},
    xtick=data,
    label style={font=\footnotesize},
    tick label style={font=\footnotesize},
    ymin=0,
    ymax=10,
    ytick={0,2,4,6,8,10},
    scaled y ticks=false,
    ymajorgrids=true,
    grid style={dashed, gray!40},
]

\addplot[fill=tabblue, draw=black,
    error bars/y dir=both, error bars/y explicit]
    coordinates {
        (1, 3.85) += (0, 4.51) -= (0, 3.85)
        (4, 4.12) += (0, 0.86) -= (0, 0.82)
    };

\addplot[fill=tabbluelight, draw=black,
    postaction={pattern=horizontal lines, pattern color=tabblue!50}]
    coordinates {
        (1, 3.87)
        (4, 5.77)
    };

\addplot[fill=taborange, draw=black,
    error bars/y dir=both, error bars/y explicit]
    coordinates {
        (1, 3.50) += (0, 0.88) -= (0, 0.82)
        (4, 3.87) += (0, 2.11) -= (0, 2.11)
    };

\addplot[fill=taborangelight, draw=black,
    postaction={pattern=horizontal lines, pattern color=taborange!50}]
    coordinates {
        (1, 3.52)
        (4, 4.48)
    };

\end{axis}
\end{tikzpicture}%
        \caption{Git Status}
    \end{subfigure}
    \begin{subfigure}{0.24\textwidth}
        \begin{tikzpicture}
\begin{axis}[
    ybar=0.4pt,
    bar width=2pt,
    width=\textwidth,
    height=0.85\textwidth,
    enlarge x limits=0.1,
    symbolic x coords={1,2,4,8,16,32,64,128},
    xtick=data,
    label style={font=\footnotesize},
    tick label style={font=\footnotesize},
    x tick label style={font=\tiny},
    ymin=0,
    ymax=4,
    ytick={0,1,2,3,4},
    scaled y ticks=false,
    ymajorgrids=true,
    grid style={dashed, gray!40},
]

\addplot[fill=tabblue, draw=black,
    error bars/y dir=both, error bars/y explicit]
    coordinates {
        (1, 0.167) +- (0, 0.006)
        (2, 0.167) +- (0, 0.006)
        (4, 0.167) +- (0, 0.006)
        (8, 0.178) +- (0, 0.006)
        (16, 0.178) +- (0, 0.006)
        (32, 0.178) +- (0, 0.006)
        (64, 0.178) +- (0, 0.006)
        (128, 0.178) +- (0, 0.006)
    };

\addplot[fill=tabbluelight, draw=black,
    postaction={pattern=horizontal lines, pattern color=tabblue!50}]
    coordinates {
        (1, 0.178)
        (2, 0.178)
        (4, 0.178)
        (8, 0.188)
        (16, 0.188)
        (32, 0.188)
        (64, 0.188)
        (128, 0.188)
    };

\addplot[fill=taborange, draw=black,
    error bars/y dir=both, error bars/y explicit]
    coordinates {
        (1, 0.162) += (0, 0.016) -= (0, 0.006)
        (2, 0.162) += (0, 0.026) -= (0, 0.006)
        (4, 0.178) += (0, 0.037) -= (0, 0.016)
        (8, 0.337) += (0, 0.483) -= (0, 0.337)
        (16, 0.236) += (0, 0.069) -= (0, 0.058)
        (32, 0.241) += (0, 0.029) -= (0, 0.013)
        (64, 0.220) += (0, 0.040) -= (0, 0.021)
        (128, 0.220) += (0, 0.058) -= (0, 0.032)
    };

\addplot[fill=taborangelight, draw=black,
    postaction={pattern=horizontal lines, pattern color=taborange!50}]
    coordinates {
        (1, 0.183)
        (2, 0.194)
        (4, 0.231)
        (8, 3.515)
        (16, 0.963)
        (32, 1.371)
        (64, 1.005)
        (128, 1.048)
    };

\end{axis}
\end{tikzpicture}%
        \caption{Conda Activate}
    \end{subfigure}
    \begin{subfigure}{0.24\textwidth}
        \begin{tikzpicture}
\begin{axis}[
    width=\textwidth,
    height=0.85\textwidth,
    xmin=-1, xmax=1,
    xtick={-0.4,0.4},
    xticklabels={\LustreSystem, \NASSystem},
    label style={font=\footnotesize},
    tick label style={font=\footnotesize},
    x tick label style={font=\tiny},
    ymin=0,
    ymax=100,
    ytick={0,20,40,60,80,100},
    scaled y ticks=false,
    ymajorgrids=true,
    grid style={dashed, gray!40},
]

\addplot[ybar, bar width=20pt, fill=tabblue, draw=black,
    error bars/y dir=both, error bars/y explicit]
    coordinates {
        (-0.4, 74.7) +- (0, 2.7)
    };

\addplot[ybar, bar width=20pt, fill=taborange, draw=black,
    error bars/y dir=both, error bars/y explicit]
    coordinates {
        (0.4, 50.7) +- (0, 8.0)
    };

\end{axis}
\end{tikzpicture}%
        \caption{Conda Create}
    \end{subfigure}
    \\[0.3em]
    \begin{tikzpicture}
    \draw[fill=tabblue, draw=black] (0,0) rectangle (0.3,0.2);
    \node[right, font=\small] at (0.35,0.1) {\texttt{LUSTRE\_SYS} (Mean)};
    \draw[fill=tabbluelight, draw=black, postaction={pattern=horizontal lines, pattern color=tabblue!50}] (3.8,0) rectangle (4.1,0.2);
    \node[right, font=\small] at (4.15,0.1) {\texttt{LUSTRE\_SYS} (P99)};
    \draw[fill=taborange, draw=black] (7.6,0) rectangle (7.9,0.2);
    \node[right, font=\small] at (7.95,0.1) {\texttt{NAS\_SYS} (Mean)};
    \draw[fill=taborangelight, draw=black, postaction={pattern=horizontal lines, pattern color=taborange!50}] (11,0) rectangle (11.3,0.2);
    \node[right, font=\small] at (11.35,0.1) {\texttt{NAS\_SYS} (P99)};
\end{tikzpicture}
    \caption{Developer-experience benchmarks on \LustreSystem{} and \NASSystem{} using \sysname{}. (a)~\texttt{git\_clone} and (b)~\texttt{git\_status} (first scan), (c)~\texttt{conda\_activate}, and (d)~\texttt{conda\_create} from an offline explicit lock file with package artifacts staged locally before timing. 
   $x$-axis ((a)--(c)): task count, $y$-axis: latency in milliseconds. Error bars: 95\% CI.}
    \label{fig:lustre_nas_devx}
\end{figure*}

AI research involves frequent source-code and environment-management operations. We use Git clone and status, and Conda create and activate, to represent these developer workflows. As shown in Figure~\ref{fig:lustre_nas_devx}, we observe that \LustreSystem{} completes Git clone 1.64--1.75~$\times$ faster than \NASSystem{}, while Git status performance is comparable and varies with client count. 
 
For Conda, \NASSystem{} reduces single-rank environment creation latency by 32\%,
whereas \LustreSystem{} provides more stable activation latency at scale; at 128 ranks, the mean per-rank P99 is 193~ms on \LustreSystem{} compared with 1.06~s on \NASSystem{}. These results highlight that researcher productivity depends on metadata performance and tail latency in addition to peak I/O throughput, and that no single storage system performs best across all workflow phases.

\section{Related Work}
\label{sec:related}

\begin{table*}[t]
\centering
\caption{Comparison of \sysname{} with prior storage benchmarks, categorized by execution model, workload awareness, and target use case. \sysname{} distinguishes itself by executing real research operations---PyTorch checkpointing and data loading, environment setup, version control, and metadata microbenchmarks---on standard POSIX filesystems.}
\label{tab:comparison}
\resizebox{\textwidth}{!}{
\begin{tabular}{l|c|c|c|c|c|c|c}
\toprule
\textbf{Dimension} & \textbf{fio}~\cite{fio} & \textbf{IOR}~\cite{ior} & \textbf{Filebench}~\cite{filebench} & \textbf{MLPerf Storage}~\cite{mlperfstorage} & \textbf{elbencho}~\cite{elbencho} & \textbf{DLIO}~\cite{dlio} & \textbf{\sysname{}} \\
\midrule
Execution Model & Synthetic & Synthetic & Synthetic & Simulated & Synthetic & Simulated & \textbf{AI Training Frameworks} \\
ML Workload Aware & \texttimes & \texttimes & \texttimes & \checkmark & \texttimes & \checkmark & \textbf{\checkmark} \\
Checkpointing & \texttimes & \texttimes & \texttimes & Partial\textsuperscript{\textdagger} & \texttimes & \checkmark\textsuperscript{\textdagger} & \textbf{\checkmark} \\
Data Loading & \texttimes & \texttimes & \texttimes & \checkmark & \texttimes & \checkmark & \textbf{\checkmark} \\
Env Setup / Metadata & \texttimes & \texttimes & Partial & \texttimes & \texttimes & \texttimes & \textbf{\checkmark} \\
POSIX Interface & \checkmark & \checkmark & \checkmark & \checkmark & \checkmark & \checkmark & \textbf{\checkmark} \\
Distributed Execution & \checkmark & \checkmark & \texttimes & \checkmark & \checkmark & \checkmark & \textbf{\checkmark} \\
Hugging Face Integration & \texttimes & \texttimes & \texttimes & \texttimes & \textbf{\checkmark} \\
Target Use Case & Raw Perf. & HPC I/O & Workloads & Prod. ML & Hybrid & Simulated ML & \textbf{AI Research} \\
\bottomrule
\multicolumn{8}{l}{\textsuperscript{\textdagger}\textit{Emulated via configured I/O patterns; does not execute PyTorch checkpoint code paths.}} \\
\end{tabular}
}
\end{table*}

Storage benchmarking has evolved over several decades, with tools designed for diverse objectives ranging from raw device characterization to application-specific performance evaluation. 
We organize and summarize related work (Table~\ref{tab:comparison}) into two primary categories: general-purpose POSIX storage benchmarks that excel at measuring peak performance but lack awareness of ML workload characteristics, and ML-specific data management systems that optimize for training workflows but sacrifice POSIX generality. 
\sysname{} takes a third path: ML-aware benchmarks that execute real operations on general-purpose POSIX filesystems, prioritizing fidelity over the convenience of simulation.



\textbf{Synthetic Microbenchmarks} like \textbf{fio}~\cite{fio}, \textbf{IOzone}~\cite{iozone}, and \textbf{Bonnie++}~\cite{bonnie} are the de facto standards for storage benchmarking. 
They provide extensive configurability for block-level I/O patterns, excelling at characterizing raw storage performance by stressing specific subsystems. However, they generate independent I/O streams that do not capture the correlated, bursty access patterns of distributed training, where storage operations are synchronized with gradient computation and collective communication. 
While these tools provide invaluable baseline measurements, the gap between their synthetic I/O and actual training workloads can lead to misleading capacity planning decisions.

\textbf{HPC Parallel Benchmarks} such as \textbf{IOR}~\cite{ior} and \textbf{mdtest}~\cite{mdtest} target large-scale parallel filesystems, measuring aggregate bandwidth and metadata operation rates across distributed processes. 
IOR coordinates file-per-process and shared-file access patterns representative of scientific simulation checkpointing, while mdtest stress-tests directory operations for job staging. 
The \textbf{IO500}~\cite{io500} benchmark combines these two to provide a holistic ranking of HPC storage systems. 
These tools assume relatively uniform access patterns across processes, which diverges from the heterogeneous I/O behavior of AI training where checkpoint writes are synchronous barriers and data loading is stochastic.

\textbf{Workload-Driven Simulators} like \textbf{Filebench}~\cite{filebench} introduce workload modeling through a domain-specific language to describe file operations and access probabilities. Users can compose synthetic workloads mimicking database transactions or web servers. However, creating accurate ML workload models requires deep understanding of framework internals and data formats---knowledge that quickly becomes outdated. \sysname{} addresses this by executing actual PyTorch operations rather than modeling them.

\textbf{Modern Hybrid Benchmarks} like \textbf{elbencho}~\cite{elbencho} provide support for distributed execution, GPU memory as a source/sink, and unified interfaces across files, objects, and block devices.
However, elbencho shares the fundamental limitation of synthetic I/O: it measures what storage \emph{can} deliver rather than what AI research \emph{will} demand. The distinction matters because research workloads exhibit complex temporal patterns that synthetic steady-state benchmarks cannot capture.



\textbf{ML-Specific Benchmarks:} \textbf{MLPerf Storage}~\cite{mlperfstorage} is the most direct effort to benchmark storage for ML workloads. It replays data-ingestion patterns at rates derived from accelerator computation speeds, running the simulation without GPUs, and its v2.0 release added checkpointing workloads. \textbf{DLIO}~\cite{dlio}, which also serves as MLPerf Storage's simulation engine, is a configurable I/O benchmark that emulates training epochs and checkpoint phases, including LLM checkpointing. The key limitation of both is that they replay configured I/O patterns rather than executing framework code: no \texttt{torch.save}, no FSDP sharded-checkpoint coordination, no genuine checkpoint file-creation metadata storms. 
They capture the \emph{shape} of training I/O but approximate the framework-specific behaviors that determine how that I/O stresses a filesystem. Neither covers environment setup, version control, or the interactive metadata diversity of research workflows.



\textbf{Object Storage Systems}, often benchmarked with tools like \textbf{COSBench}~\cite{cosbench} and \textbf{warp}~\cite{warp}, provide high-throughput sample serving but represent a different architectural choice. Replacing POSIX with purpose-built abstractions requires significant infrastructure investment and does not address the broader storage needs of research clusters, including code repositories and experiment artifacts.

\paragraph{Positioning \sysname{}}

\sysname{} differs from the benchmarks surveyed above in one basic respect: it executes rather than emulates. It is a lock-stock-and-barrel suite that runs on the cluster's existing CPU and GPU nodes and performs the operations researchers actually use: PyTorch DDP and FSDP checkpointing across a range of Hugging Face architectures, distributed data loading through framework data loaders, version-control and environment-setup commands, and metadata microbenchmarks, all against the filesystem under test. Because the benchmarks run real framework code paths, they track framework behavior as it evolves; captured I/O patterns go stale, and configured emulations omit behaviors such as checkpoint coordination and file-creation metadata dynamics.

That difference matters in practice. A filesystem can post excellent fio numbers while checkpointing poorly under real framework behavior, for reasons such as metadata overhead or lock contention that synthetic I/O does not reproduce. Our regression study is a case in point: it surfaced an \texttt{openat} lock-contention pathology under FSDP checkpointing that pattern replay would not have produced (\S~\ref{sec:regression}). \sysname{} packages this approach as a reusable methodology for AI research clusters: a systematic set of research-relevant storage operations, spanning research development environments, data loading, and checkpointing, in an extensible framework, runnable wherever researchers already run.

\section{Conclusion}
The performance of storage systems is a critical, yet often overlooked, factor in the productivity of AI research clusters. 
The dynamic, metadata intensive workloads common in research differ significantly from predictable, throughput oriented production training patterns. 
This divergence means traditional benchmarks fail to adequately model real world usage, leading to suboptimal infrastructure choices. 
To address this, we developed \sysname{}, a benchmark suite that provides a high fidelity assessment of storage performance by executing real PyTorch operations that mirror the complete AI research lifecycle.

We demonstrate \sysname{}'s significant impact as a validation and benchmarking suite for the specialized storage needs of AI research. 
Its utility is proven in establishing performance baselines for new cluster deployments, enabling objective architectural comparisons that inform workload aware provisioning, and detecting performance regressions as a continuous integration tool. 
By capturing these complex, real world behaviors, \sysname{} provides actionable insights that emulated synthetic benchmarks cannot.

In conclusion, \sysname{} bridges the gap between generalized storage benchmarking and the specific demands of AI research. 
It offers a robust, data driven methodology for infrastructure teams and vendors to select, validate, and optimize storage solutions. 
By ensuring storage is purpose fit for its demanding role, \sysname{} empowers organizations to maximize GPU cluster efficiency, accelerate innovation, and realize the full potential of their research investments.

\clearpage
\newpage
\bibliographystyle{assets/plainnat}
\bibliography{refs}

\begin{thebibliography}{12}
\providecommand{\natexlab}[1]{#1}
\providecommand{\url}[1]{\texttt{#1}}
\expandafter\ifx\csname urlstyle\endcsname\relax
  \providecommand{\doi}[1]{doi: #1}\else
  \providecommand{\doi}{doi: \begingroup \urlstyle{rm}\Url}\fi

\bibitem[Axboe(2023)]{fio}
Jens Axboe.
\newblock fio - flexible {I/O} tester.
\newblock \url{https://github.com/axboe/fio}, 2023.
\newblock Accessed: 2025-01-28.

\bibitem[Breuner(2023)]{elbencho}
Sven Breuner.
\newblock elbencho: A distributed storage benchmark for files, objects and blocks with support for {GPUs}.
\newblock \url{https://github.com/breuner/elbencho}, 2023.
\newblock Accessed: 2025-01-28.

\bibitem[Coker(2023)]{bonnie}
Russell Coker.
\newblock Bonnie++ benchmark suite.
\newblock \url{https://www.coker.com.au/bonnie++/}, 2023.
\newblock Accessed: 2025-01-28.

\bibitem[Devarajan et~al.(2021)Devarajan, Zheng, Kougkas, Sun, and Vishwanath]{dlio}
Hariharan Devarajan, Huihuo Zheng, Anthony Kougkas, Xian-He Sun, and Venkatram Vishwanath.
\newblock {DLIO}: A data-centric benchmark for scientific deep learning applications.
\newblock In \emph{Proceedings of the IEEE/ACM 21st International Symposium on Cluster, Cloud and Internet Computing (CCGrid)}, pages 81--91. IEEE, 2021.

\bibitem[Kunkel et~al.(2017)Kunkel, Dilger, and Breuner]{io500}
Julian~M. Kunkel, Andreas Dilger, and Sven Breuner.
\newblock Establishing the {IO-500} benchmark.
\newblock In \emph{Proceedings of the Workshop on Performance and Scalability of Storage Systems (PDSW-DISCS)}. Virtual Institute for I/O, 2017.

\bibitem[{Lawrence Livermore National Laboratory}(2023)]{mdtest}
{Lawrence Livermore National Laboratory}.
\newblock mdtest: Metadata performance benchmark.
\newblock \url{https://github.com/hpc/ior}, 2023.
\newblock Part of the IOR repository.

\bibitem[{MinIO, Inc.}(2023)]{warp}
{MinIO, Inc.}
\newblock warp: {S3} benchmarking tool.
\newblock \url{https://github.com/minio/warp}, 2023.
\newblock Accessed: 2025-01-28.

\bibitem[{MLCommons}(2025)]{mlperfstorage}
{MLCommons}.
\newblock {MLPerf} storage benchmark suite.
\newblock \url{https://github.com/mlcommons/storage}, 2025.
\newblock Version 2.0.

\bibitem[Norcott and Capps(2016)]{iozone}
William Norcott and Don Capps.
\newblock {IOzone} filesystem benchmark.
\newblock \url{https://www.iozone.org/}, 2016.
\newblock Accessed: 2025-01-28.

\bibitem[Shan and Shalf(2007)]{ior}
Hongzhang Shan and John Shalf.
\newblock Using {IOR} to analyze the {I/O} performance for {HPC} platforms.
\newblock In \emph{Proceedings of the Cray User Group}. Lawrence Berkeley National Laboratory, 2007.

\bibitem[Tarasov et~al.(2016)Tarasov, Zadok, and Shepler]{filebench}
Vasily Tarasov, Erez Zadok, and Spencer Shepler.
\newblock Filebench: A flexible framework for file system benchmarking.
\newblock \emph{USENIX ;login:}, 41\penalty0 (1):\penalty0 6--12, 2016.

\bibitem[Zheng et~al.(2013)Zheng, Chen, Wang, Zhang, and Duan]{cosbench}
Qingsong Zheng, Hai Chen, Yongjun Wang, Jianhua Zhang, and Jie Duan.
\newblock {COSBench}: Cloud object storage benchmark.
\newblock In \emph{Proceedings of the 4th ACM/SPEC International Conference on Performance Engineering (ICPE)}, pages 199--210. ACM, 2013.

\end{thebibliography}

\clearpage
\newpage
\beginappendix

\section{Acknowledgments}

We would like to thank Shubho Sengupta, Kim Hazelwood, Vivek Pai, Kevin Lee for their pioneering work on 
architecting and building high performance research clusters with a focus on usability for FAIR. Building successive versions of the clusters with varying POSIX storage solutions, 
shaped the vision for \sysname{}, from its origins as a simple file-creation validation tool to a benchmark suite tailored for AI research storage workloads. 
We are also grateful to the Cloud Foundation team (Chris Henry, Chandan Avdhut, Kareem Sakher, Christopher Sherman, and Wes Miravate) at Meta for their generous support in standing up the filesystems and provisioning the compute clusters that made this work possible.

\end{document}